\newcommand{\HII}{H\,{\tiny II}}
\newcommand{\hii}{H\,{\footnotesize II}}

\newcommand{\CII}{[C\,{\tiny II}]}

\newcommand{\HIIR}{\HII R}
\newcommand{\cHIIR}{c\HIIR}

\newcommand{\twCII}{[$^{\rm 12}$C\,{\tiny II}]}
\newcommand{\thCII}{[$^{\rm 13}$C\,{\tiny II}]}

\newcommand{\OI}{[O\,{\footnotesize I}]}

\newcommand{\kms}{\mathrm{km\,s^{-1}}}
\newcommand{\kkms}{\mathrm{K\,km\,s^{-1}}}

\newcommand{\mum}{\mathrm{\mu m}}

\newcommand{\e}{{\rm e}}

\newcommand{\Tex}{T_\mathrm{ex}}

\documentclass[letter]{aa}
\usepackage{graphicx}
\usepackage{txfonts}
\usepackage[colorlinks=true,     linkcolor=blue, citecolor=blue, filecolor=blue, urlcolor=blue]{hyperref}

\begin{document} 


\title{[C\,{\large II}]-deficit caused by self-absorption in an ionized carbon-filled bubble in RCW79}



\author{E. Keilmann\inst{1}
\and S. Dannhauer\inst{1} 
\and S. Kabanovic\inst{1}   
\and N. Schneider\inst{1}    
\and V. Ossenkopf-Okada\inst{1}     
\and R. Simon \inst{1}
\and L. Bonne\inst{2}
\and \\ P.~F. Goldsmith\inst{3}
\and R. G\"usten \inst{4} 
\and A. Zavagno \inst{5,6} 
\and J. Stutzki \inst{1}
\and D. Riechers \inst{1} 
\and M. R\"ollig \inst{7,1}    
\and \\J.~L. Verbena \inst{1} 
\and A.~G.~G.~M. Tielens \inst{8,9} 
}

\institute{I. Physikalisches Institut, Universität zu K\"{o}ln,  Z\"ulpicher Stra\ss{}e 77, 50937 K\"oln, Germany \\
      \email{keilmann@ph1.uni-koeln.de}
\and{SOFIA Science Center, USRA, NASA Ames Research Center, Moffett Field, CA 94 045, USA} 
\and{Jet Propulsion Laboratory, California Institute of Technology, 4800 Oak Grove Drive, Pasadena, CA 91109-8099, USA} 
\and{Max-Planck Institut f\"ur Radioastronomie, Auf dem H\"ugel 69, 53121 Bonn, Germany} 
\and{Aix Marseille Univ, CNRS, CNES, LAM, Marseille, France}  
\and{Institut Universitaire de France, 1 rue Descartes, Paris, France} 
\and Physikalischer Verein, Gesellschaft für Bildung und Wissenschaft, Robert-Mayer-Str. 2, 60325 Frankfurt, Germany 
\and{University of Maryland, Department of Astronomy, College Park, MD 20742-2421, USA} 
\and{Leiden Observatory, PO Box 9513, 2300 RA Leiden, The Netherlands} 
}

\date{draft of \today}

\titlerunning{A C$^+$-filled bubble in RCW79} 

\authorrunning{E. Keilmann}

\abstract 
{
Recent spectroscopic observations of the \CII\ 158$\,\mum$ fine-structure line of ionized carbon (C$^+$), using the Stratospheric Observatory for Infrared Astronomy (SOFIA), have revealed expanding \CII\ shells in Galactic \HII\ regions. 
We report the discovery of a bubble-shaped source (S144 in RCW79 in the GLIMPSE survey), associated with a compact \HII\ region, excited by a single O7.5--9.5V/III star, which is consistent with a scenario that the bubble is still mostly ``filled'' with C$^+$. This indicates most likely a very early evolutionary state, in which the stellar wind has not yet blown material away as it has in more evolved \HII\ regions.  
Using the SimLine non-local thermodynamic equilibrium radiative transfer code, the \CII\ emission can be modeled to originate from three regions: first, a central \HII\ region with little C$^+$ in the fully ionized phase, followed by two layers with a gas density around $2500\,\mathrm{cm^{-3}}$ of partially photodissociated gas. From these two layers, the second layer is a slowly expanding \CII\ shell with an expansion velocity of $\sim\,$$2.6\,\kms$ that corresponds approximately to a bright ring at 8$\,\mum$. The outermost layer exhibits a temperature and velocity gradient that produces the observed self-absorption features in the optically thick \CII\ line ($\tau \sim 4$), leading to an apparent deficit in \CII\ emission and a low ratio of \CII\ to total far-infrared (FIR) emission. 
We developed a procedure to reconstruct the missing \CII\ flux and find a linear correlation between \CII\ and FIR without a \CII-deficit after incorporating the missing \CII\ flux. This example demonstrates that at least some of the \CII-deficit found in Galactic \HII\ bubbles can be attributed to self-absorption, although a broader sample of these objects needs to be studied to fully constrain the range of conditions in which \CII-deficits can be explained by this process.
} 

\maketitle
%

\begin{figure}[htbp]
  \centering
  \includegraphics[width=0.85\linewidth]{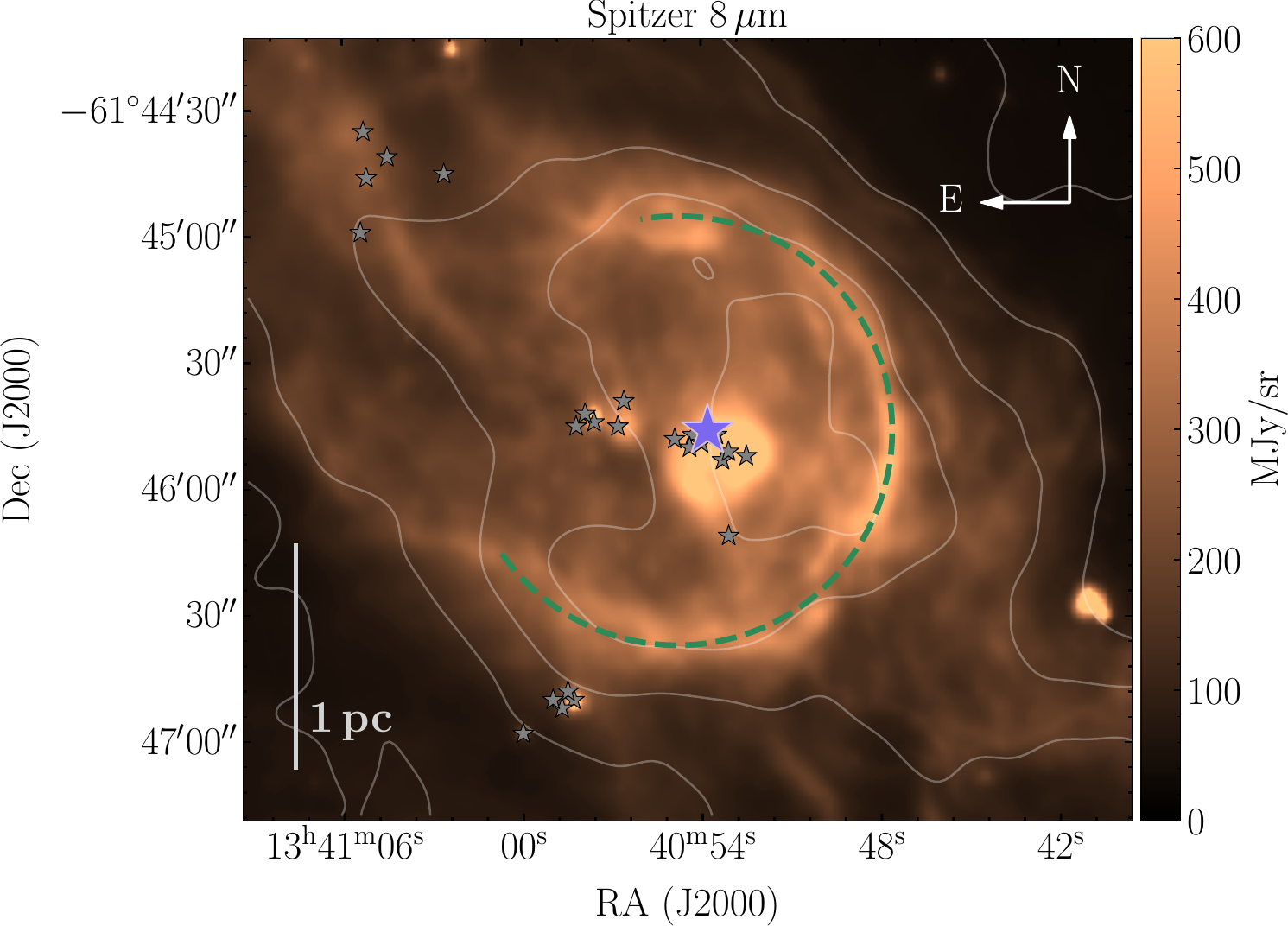}
  \caption
      {\textit{Spitzer} $8\,\mum$ image of the bubble at $\sim\,$$2''$ angular resolution from the GLIMPSE survey~\citep{Churchwell2006}. The blue star is the O7.5--9.5V/III star and the small black stars are members of the associated IR clusters. The green circle is a by-eye approximation of the bright IR ring.
      The contour lines show \CII\ emission from 50 to $250\,\kkms$ in steps of $50\,\kkms$. 
      }
    \label{fig:spitzer}
\end{figure}
\begin{figure}[htbp]
  \centering
  \includegraphics[width=0.80\linewidth]{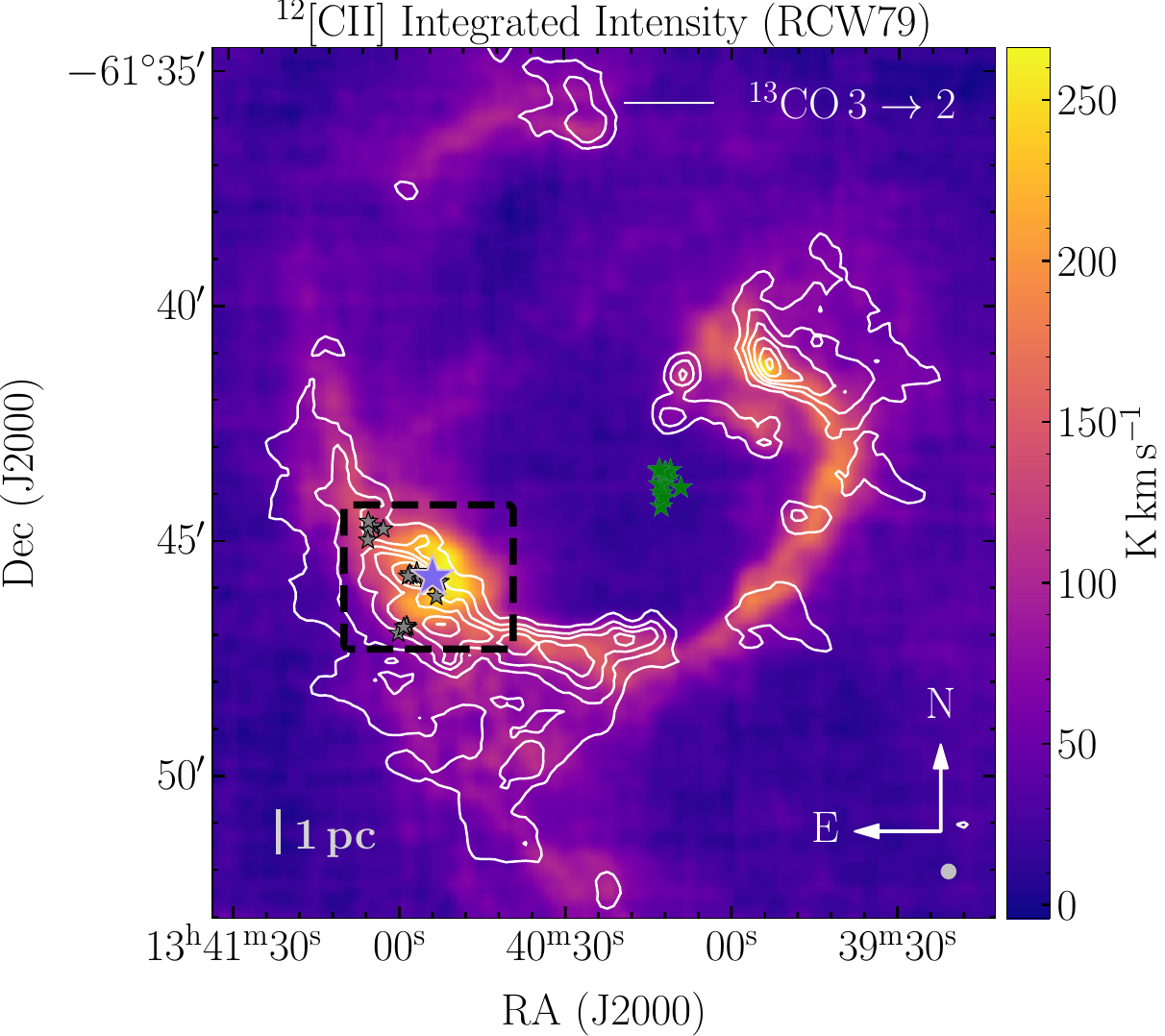}
  \includegraphics[width=0.80\linewidth]{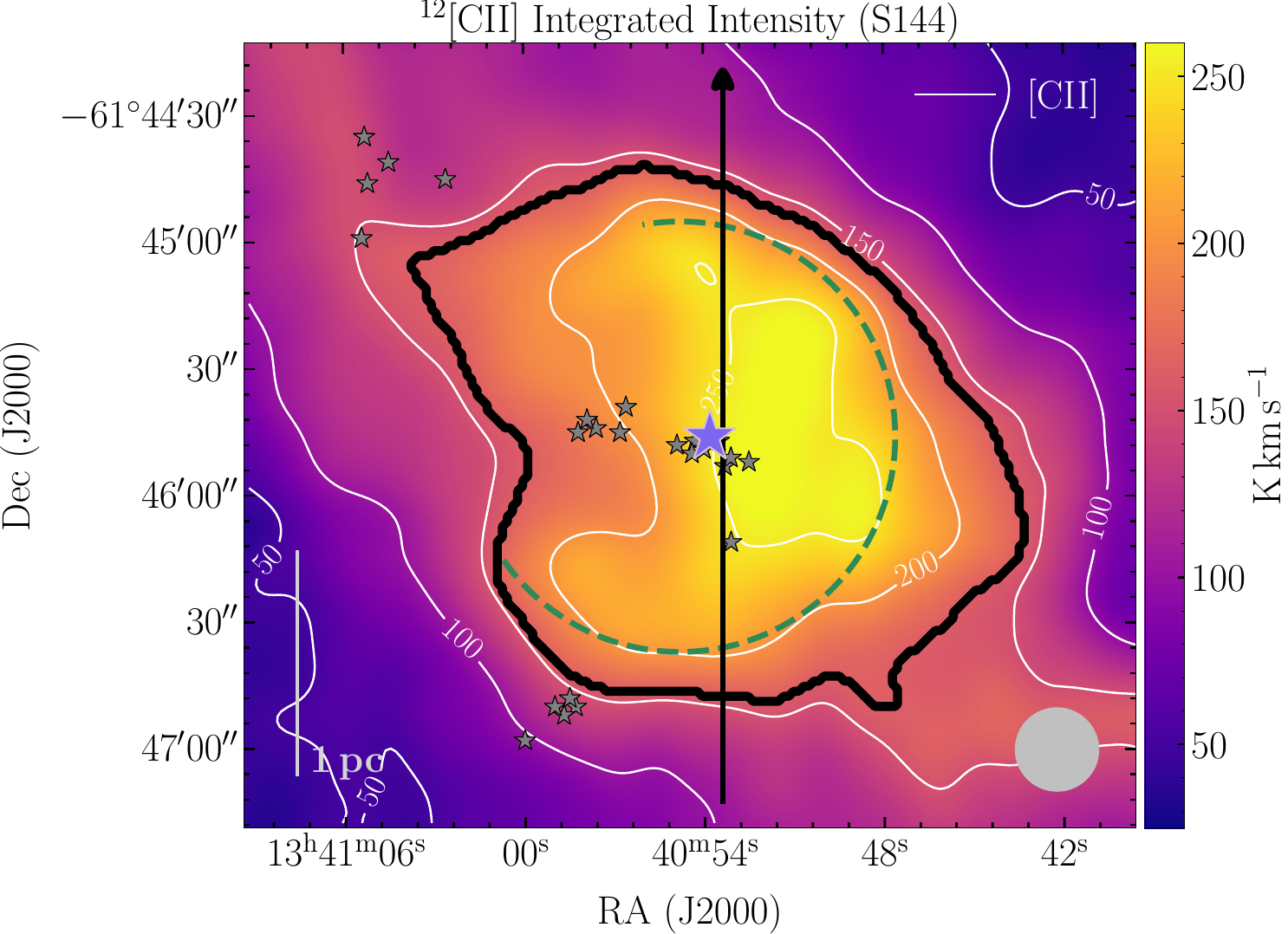}
  \caption
      {Line-integrated ($-70$ to $-20\,\kms$) \CII\ intensity maps of RCW79 and S144. Top: RCW79 with an overlay of $^{13}\mathrm{CO\,}$3$\,\to\,$2 emission with contours from 7 to $52\,\kkms$ in steps of $9\,\kkms$ ($\sim\,$$5\sigma$). The green stellar symbols indicate the central cluster of O-type stars~\citep{Martins2010}, the blue stellar symbol the exciting O7.5--9.5V/III star of the \cHIIR\ within S144 in the southeast, and the small black stellar symbols members of the  IR clusters~\citep{Zavagno2006}. The dashed black rectangle outlines the area shown in the bottom panel.
      Bottom: S144 with an overlay of $^{13}\mathrm{CO\,}$3$\,\to\,$2 at the same levels as in the upper panel. The dashed green circle ($50''$ radius) indicates the bright IR ring seen in the \textit{Spitzer} $8\,\mum$ map (see Fig.~\ref{fig:spitzer}). The region taken into account to compute the average \CII\ spectrum is indicated by a black contour.
      The black line corresponds to the PV cut shown in Fig.~\ref{fig:CII_PVcutVertical}.
      Both maps have an angular resolution of $20''$, indicated by gray disks in the lower right corner of the panels.}
    \label{fig:cii_integrated_map}
\end{figure}

\section{Introduction} \label{sec:intro}


Recent spectroscopic studies of Galactic \HII\ regions in the $158\,\mum$ line of ionized carbon (\CII) have confirmed  that this far-infrared (FIR) fine-structure line is a key tracer for the cooling processes and the dynamics of gas in photodissociation regions (PDRs). Exploratory observations have discovered expanding \CII\ shells, e.g.\@ in Orion~\citep{Pabst2019} and in FEEDBACK~\citep{Schneider2020} Galactic \HII\ region bubbles such as RCW120~\citep{Luisi2021}, using large maps of \CII\ obtained with the Stratospheric Observatory for Infrared Astronomy (SOFIA). With velocities of up to $15\,\kms$, the shell expansion is mostly attributed to be driven by stellar winds of massive stars and not to the thermal expansion of the \HII\ region.

These examples underline the potential of \CII\ to trace the evolution of an \HII\ region and its related molecular cloud. 
In this study, we report the detection of a C$^+$ bubble in a very early evolutionary phase, designated as S144, in the  \textit{Spitzer}/GLIMPSE survey~\citep{Churchwell2006}. The $8\,\mum$ emission (Fig.~\ref{fig:spitzer}) exhibits a bright infrared (IR) ring with an opening in the northeast (NE) and moderate emission inside the ring, except for a concentrated peak around the exciting O7.5--9.5V/III star. S144 is embedded in the southeastern (SE) PDR ring outside of the larger \HII\ region RCW79. 
Figure~\ref{fig:cii_integrated_map} (top) presents the \CII\ emission from the PDR ring of RCW79 and the associated molecular cloud fragments~\citep{Bonne2023}. The bottom panel illustrates that the \CII\ (as the 8 $\mu$m) emission from S144 does not exhibit a central void.

The diameter of S144 (Fig.~\ref{fig:cii_integrated_map}) is about $1.5'$, corresponding to $1.7\,$pc at a distance of $3.9\,$kpc~\citep{Bonne2023}. S144 is deeply embedded in a massive molecular clump (``condensation 2,''~\citealt{Zavagno2006}) with a mass of $\sim\,$$5000\,\mathrm{M_\odot}$~\citep{Liu2017}, 
excited by an O7.5--9.5V/III star~\citep{Martins2010}. 
The molecular gas bulk emission velocity ranges between $-50$ and $-44\,\kms$.
The region is undergoing massive star formation, witnessed by a small cluster of IR sources~\citep{Zavagno2006}. The source is classified as a compact \HII\ region (\cHIIR) with a determined $5\,$GHz total flux of $1\,$Jy~\citep{Zavagno2006}, corresponding to an ionizing flux of $1.9\times10^{48}$ photons $\mathrm{s^{-1}}$, typical of an O8 star.
Assuming spherical symmetry for the \cHIIR\ region, the authors calculated a dynamical age of $0.13\,$Myr.

We have studied the gas dynamics and excitation conditions of photodissociated gas in the S144 bubble to understand the evolution of \HII\ regions and their molecular cloud. We show that the low observed ratio of \CII\ to the total FIR luminosity, the so-called \CII-deficit~\citep{Smith2017, Pabst2021}, can be explained by \CII\ self-absorption.

\section{Observations} \label{sec:obs}

\subsection{SOFIA} \label{subsec:sofia}
The \twCII\ line at $157.74\,\mum$ was mapped across RCW79 ($\sim$$\,470\,\mathrm{arcmin^2}$) in the on-the-fly (OTF) mode, using the upgraded German REceiver for Astronomy at Terahertz frequencies~\citep[upGREAT;][]{Risacher2018} on board SOFIA. 
A forward efficiency of $\eta_\mathrm{f} = 0.97$ and main beam efficiencies, $\eta_\mathrm{mb}$, between 0.63 and 0.69 were applied to 
obtain main beam temperatures (for further observational information and details on the SOFIA legacy program FEEDBACK, see~\citealt{Schneider2020} and~\citealt{Bonne2023}). The generic angular resolution is $14.1''$ and the velocity resolution was binned to $0.3\,\kms$ wide channels. 

\subsection{APEX} \label{subsec:apex}
The $^{12}\mathrm{CO\,}$ and $^{13}\mathrm{CO\,}$3$\,\to\,$2 transitions 
at $345.80\,$GHz and $330.59\,$GHz were mapped in OTF mode with the LAsMA receiver on the APEX telescope and presented in~\citet{Bonne2023}. The data, in main beam temperature units using $\eta_\mathrm{mb} = 0.68$ with a velocity resolution of $0.3\,\mathrm{km\,s^{-1}}$, had a first-order baseline removed from the spectra at $18''$ resolution. 
In May 2024, a $180'' \times 180''$ $^{12}$CO 6$\,\to\,$5 ($\nu=691.47\,$GHz) map centered on S144 was observed in OTF mode with the SEPIA660 receiver~\citep{Belitsky2018}. The data have an angular resolution of $9''$ and a velocity resolution of $0.25\,\mathrm{km\,s^{-1}}$, leading to a root-mean-square noise level of $0.59\,$K. Spectra were processed by removing a third-order spectral baseline and applying a main beam efficiency of $\eta_\mathrm{mb} = 0.6$.

\section{Results} \label{sec:results}

\begin{figure}[htbp]
  \centering
  \includegraphics[width=0.76\linewidth]{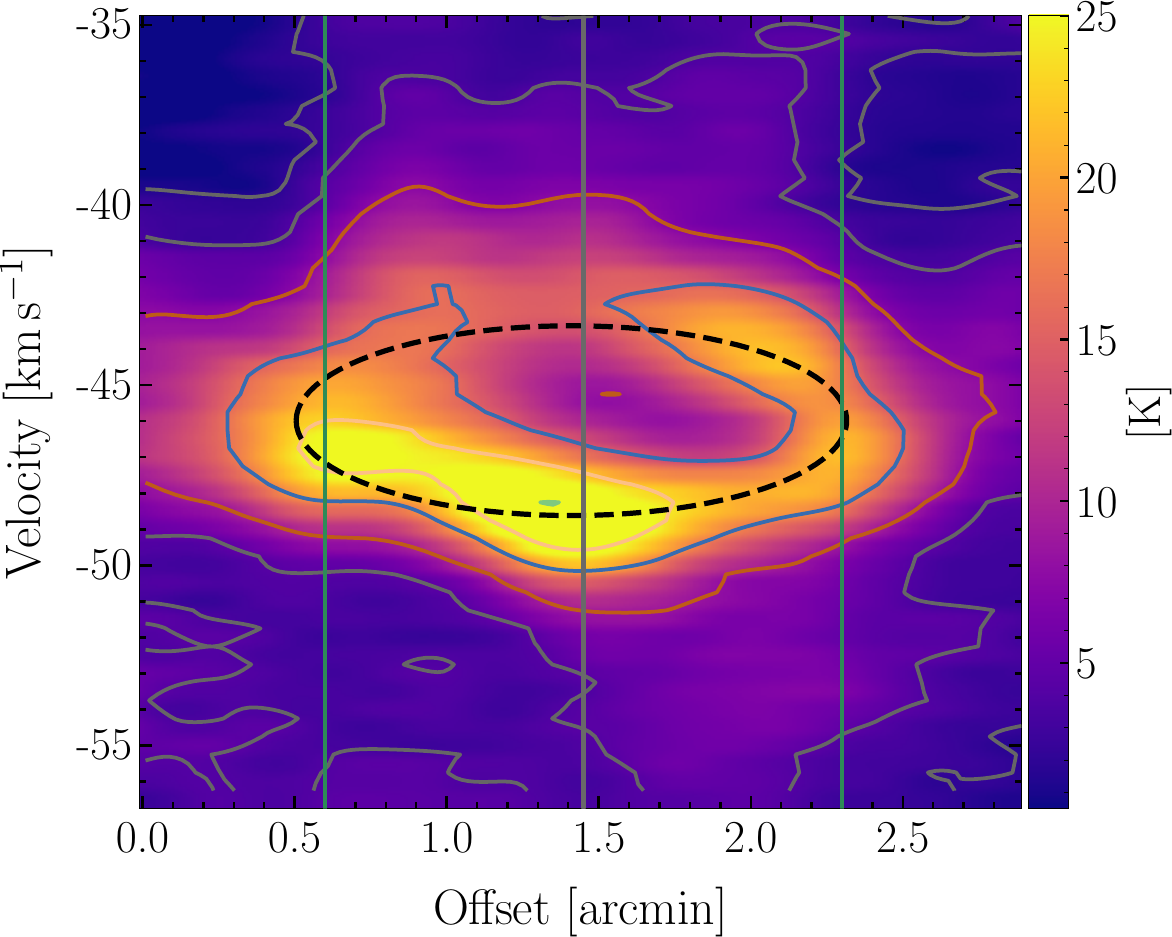}
  \caption
     {Position-velocity cut in \CII\ at 20$''$ angular resolution. The cut is outlined in the lower panel in Fig.~\ref{fig:cii_integrated_map}. Offset 0 arcmin marks the lower declination. The dashed ellipse results from fitting the emission in various PV cuts. Movies are available \href{https://hera.ph1.uni-koeln.de/~nschneid/feedback.html}{online}. 
     In these PV diagrams, the dotted black line marks the cloud bulk velocity at $-46.5\,\kms$, the vertical gray line the position of the O star, and the green lines the extent of the IR ring.   
      }
    \label{fig:CII_PVcutVertical}
\end{figure}

\subsection{Spatial and kinematic distribution of \CII\ emission} \label{subsec:overview}

We observe centrally concentrated \CII\ emission with a peak slightly to the west of the exciting O star; the bottom panel of Fig.~\ref{fig:cii_integrated_map} displays the \CII\ line-integrated intensity in S144. The $8\,\mum$ image (Fig.~\ref{fig:spitzer}) shows a more complex emission distribution with a small, circular emission peak just around the O star (radius $\sim\,$$15''$), followed by low-surface-brightness emission and a bright dust ring,
which most likely represent a swept-up gas shell, as has been proposed for other \HII\ regions~\citep{Deharveng2010}. 
The 8$\,\mum$ emission is dominated by the 7.7 and 8.6$\,\mum$ features of polycyclic aromatic hydrocarbons (PAHs) that are easily destroyed by the hard radiation field of the central star or blown out by stellar winds~\citep{Churchwell2006}. The absence of a central void in the 8$\,\mum$ emission  
suggests that the \cHIIR\ region is in an early evolutionary stage in which the gas dynamics are not yet dominated by stellar winds. Additionally, the UV radiation is attenuated by gas and dust, creating conditions that allow PAHs to survive. 

Lastly, we have a scenario in which the central O star has created a \cHIIR\ but no wind-blown cavity, and in which the star is enveloped by an extended PDR seen in the (F)IR and in \CII. The whole region is embedded in a larger molecular cloud that extends mostly to the SE, containing a few dense clumps with ongoing star formation. 
Figure~\ref{fig:complementary_plots} presents overlays of \CII\ with several other tracers (\textit{Herschel} $70\,\mum$, $843\,$MHz continuum, etc.), a \CII\ and $^{12}$CO 3$\,\to\,$2 channel map (Fig.~\ref{fig:CII_channel_map_co32Contours}), and an overlay between \CII\ and $^{12}$CO 6$\,\to\,$5 emission (Fig.~\ref{fig:spectra}). The online material includes movies that further detail the bubble's characteristics.

Figure~\ref{fig:CII_PVcutVertical} presents a position-velocity (PV) cut of the \CII\ emission showing the characteristic emission distribution of an expanding shell, with both blueshifted and redshifted parts. The latter is sometimes missing in other sources. The expansion speed is low, around $2.6\,\kms$, and was derived by fitting an ellipse to the emission distributions for all PV cuts and taking the overall average. The shell most likely corresponds to the bright IR ring (Fig.~\ref{fig:spitzer}); whether it is driven by wind or thermal pressure will be discussed in another study. 
The central dip in the profiles may reflect the effects of absorption by a colder foreground gas. The emission within the dip reaches $10-15\,$K, which is above the $3\sigma$ noise ($\sim\,$$7.5\,$K in a $0.5\,\kms$ channel), indicating significant \CII\ emission at the center of the bubble. 
In the next section, we show that optical depth effects and self-absorption lead to the apparent depression in the emission line profile.

\begin{figure}[htbp]
\centering
\hspace{-0.8cm} \includegraphics[width=0.8\linewidth]{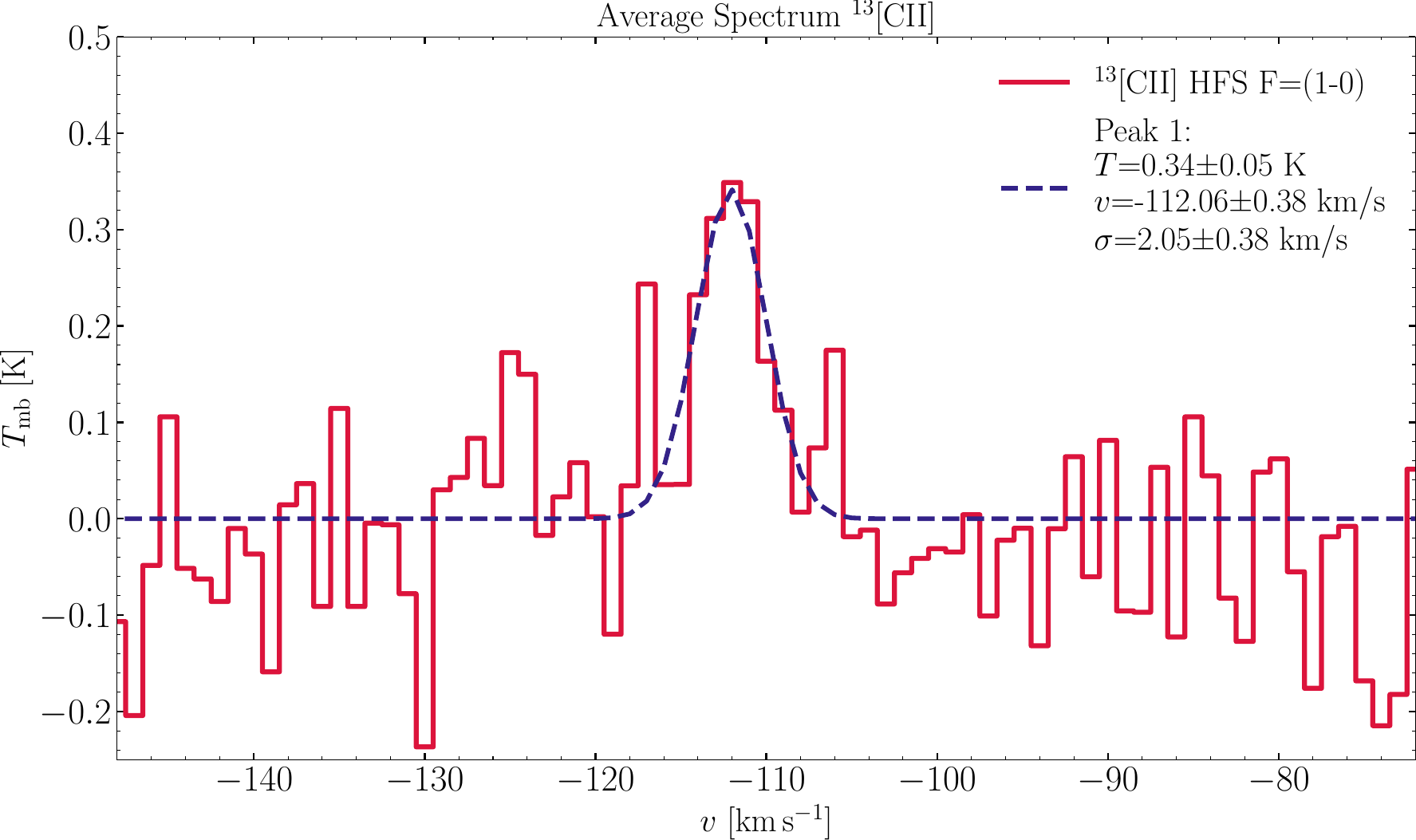}
    \centering
 \includegraphics[width=0.84\linewidth]{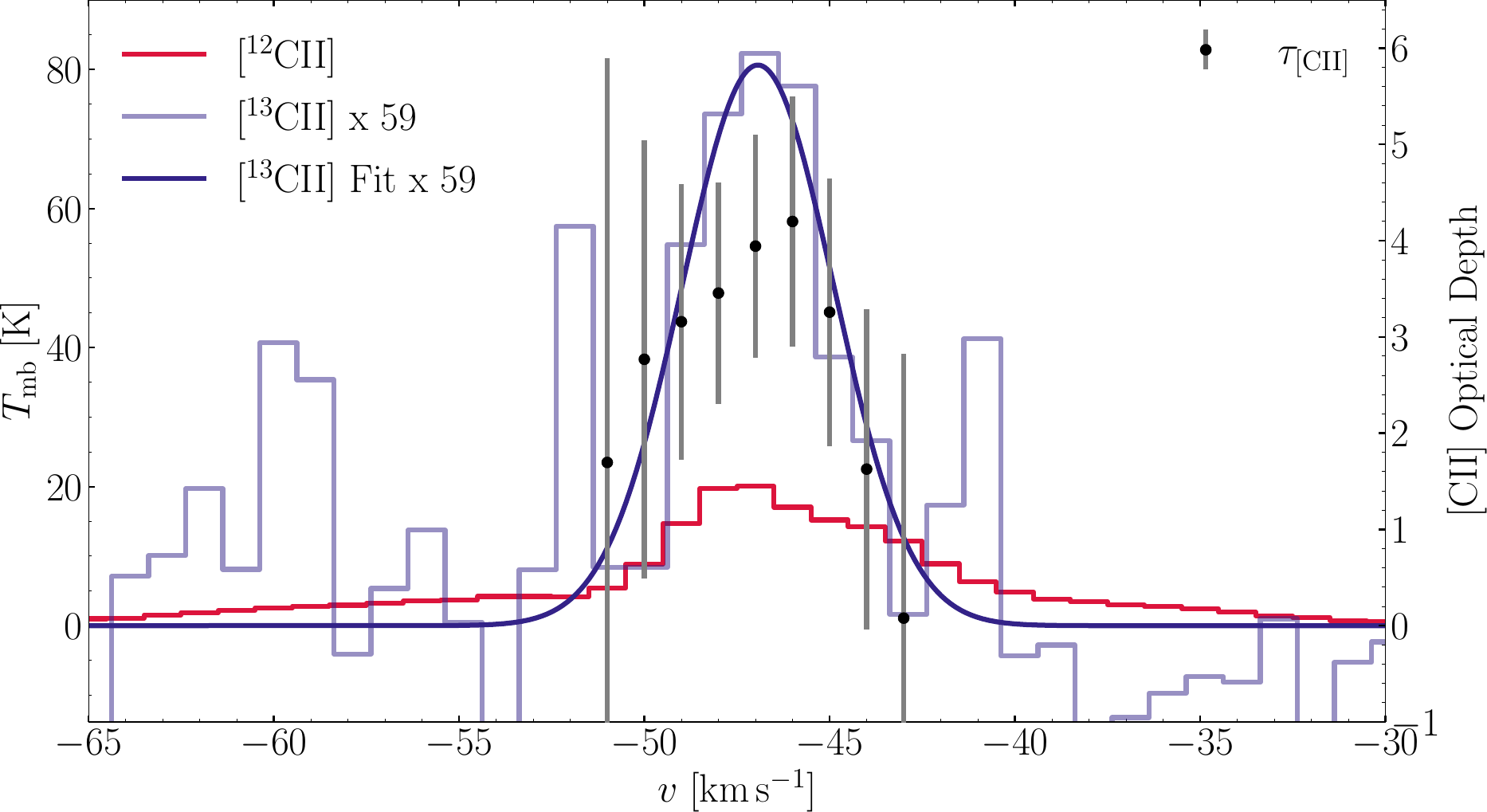}
  \caption
      {Spatially averaged \twCII\ and \thCII\ spectra. 
      Top: Averaged spectrum of the \thCII\ HFS $\mathrm{F}=1-0$ component. The spatial region over which the averaging was carried out is defined by the black dendrogram contour in Fig.~\ref{fig:cii_integrated_map}. Gaussian fit parameters are included. 
      Bottom: Averaged \twCII\ line (red) and the \thCII\ HFS $\mathrm{F}=1-0$ component (light blue histogram) with its Gaussian fit (dark blue curve), scaled by the $^{12}$C/$^{13}$C ratio of $59\pm10$~\citep{Milam2005} and considering the fractional weight (0.25) of the $1-0$ HFS. Black dots indicate optical depth.
      }
    \label{fig:CII_average_spec_opticalDepth}
\end{figure}
\subsection{Self-absorption of the \twCII\ line} 
\label{subsec:self}

S144 has a complex spatial and velocity structure, which is also reflected in the spectra of \CII\ and CO 6$\,\to\,$5 emission (Fig.~\ref{fig:spectra}). The \CII\ and CO spectra often reveal two velocity components and a dip at $\sim\,$$-46\,\kms$. In the SE region, the CO line consists of only one component at the bulk emission of $-46.5\,\kms$. In the northwestern (NW) region, CO emission decreases, while the \CII\ line is strong and shows two velocity components. In addition, the \CII\ spectrum displays in all spectra high-velocity gas from the large expanding \CII\ shell of RCW79~\citep{Bonne2023}, which is lacking in the CO line. There is also \CII\ emission arising from the slowly expanding shell of the bubble and from the bulk emission of the PDR gas, in which the bubble is embedded.

To answer the main question -- whether the \CII\ intensity dip is due to the kinematics of the expanding \CII\ shell of the bubble or due to self-absorption effects in an optically thick \CII\ line, as is seen in other C$^+$  bubbles~\citep{Bonne2022, Kabanovic2022} -- we need an observation of at least one hyperfine-structure (HFS) component of the optically thin \thCII\ line~\citep{Ossenkopf2013}. 
For S144, we do not observe this line in individual spectra. We detect the \thCII\ $\mathrm{F}=1-0$ component\footnote{The strongest \thCII\ $\mathrm{F}=2-1$ component is not accessible, as it is contaminated by the redshifted wing of the \CII\ line.} only in the averaged spectrum.
Figure~\ref{fig:CII_average_spec_opticalDepth} shows the line together with the \CII\ optical depth derived from the observed \twCII/\thCII-ratio in each velocity channel within the single-layer model
using Eq.~(\ref{eq:tau_cii}) 
(\citealt{Guevara2020},~\citealt{Kabanovic2022}, and Appendix~\ref{app:two_layer_model}). 
We find that the \thCII\ line, scaled by the local carbon abundance ratio, overshoots the \twCII\ line, resulting in a velocity-resolved optical depth between 1 and 5. The peak optical depth is shifted relative to the \CII\ peak emission, which cannot be explained if the optical depth arises purely from a gas column with a single temperature. Thus, a more sophisticated calculation is needed to properly explain the observations, separating the gas into a warm emitting layer and a colder absorbing foreground gas. The two-layer model and its results are discussed in Sect.~\ref{app:two_layer_model}.
In the following, we assume that the \twCII\ line is optically thick throughout the map, which is a presumption because we do not have measurements of the \thCII\ line at each position. However, all observed \twCII\ lines show a dip or sometimes a flat-top profile (Fig.~\ref{fig:spectra}) and considering the small extent of the region (diameter $\sim\,$$2\,$pc), we do not expect a large variation in the bulk emission of
the gas, which would shift the \twCII\ dip and the \thCII\ line center position. 

To verify our finding of partly self-absorbed \CII\ profiles, we used the 1D non-local thermodynamic equilibrium radiative transfer code SimLine~\citep{Ossenkopf2001}.
This evaluation is not intended to provide a fully quantitative reproduction of the bubble parameters, as the angular resolution of the \CII\ data imposes certain limitations.\footnote{Note, however, that the angular resolution is handled fully self-consistently within SimLine.}  
Our goal is to demonstrate that even a simple model can satisfactorily reproduce the observations. 
The model consists of a central, fully ionized \HII\ region (with a radius of $r=0.5\,$pc, electron density of $n_e=100\,\mathrm{cm^{-3}}$, temperature of $T=8000\,$K), followed by a dense ($n=2500\,\mathrm{cm^{-3}}$) PDR layer ($r=0.95\,$pc) with $T=100\,$K, surrounded by a layer in which the parameters drop to the environmental conditions reached at $r=1.5\,$pc using a power law with an exponent of $-2$. 
The PDR layer exhibits a radial velocity of $2.6\,\kms$ and a turbulent velocity of $2.2\,\kms$, consistent with observational constraints. A detailed description of the model parameters and their values is provided in Appendix~\ref{app:SimLine}.

The radius and width of the shells, the expansion velocity, and the steep outer temperature gradient are well constrained. Conversely, the local density and temperature are uncertain. A lower density can always be compensated for by a higher temperature and vice versa.  
The 8\,$\mum$ ring likely traces a compressed shell that is somewhat thinner than the modeled PDR layer. This indicates that C$^+$ can be extended more than the PAHs in S144. However, these structural differences do not impact our key conclusion: the observed \CII\ line can be self-absorbed due to temperature and velocity gradients.
The exact amount of C$^+$ contained in the \HII\ region remains currently unclear.
Modeling the \HII\ region with the assumed parameters from SimLine using the CLOUDY spectral synthesis code~\citep{Ferland2017} yields very small quantities of C$^+$ (less than 1\%), although up to 20\% of the observed \CII\ can stem from the \HII\ regions in other sources (RCW120,~\citealt{Luisi2021}). 
Observations of carbon recombination lines could probably settle this issue and help investigate the dynamics of the expanding \CII\ shell.

\subsection{\CII-deficit} \label{subsec:deficit}

The self-absorption seen in the \CII\ line leads to a deficiency in the line-integrated emission. Thus, the ratio of \CII\ to total FIR continuum is lower than the nominal value, assuming that the FIR emission is optically thin, an assumption that is reasonable given the density (a few $1000\,\mathrm{cm^{-3}}$, Sect.~\ref{subsec:self}) and the radiation field regime (up to a few $1000\,\mathrm{G_0}$; Appendix~\ref{app:complementary_data}) present in S144~\citep{Goldsmith2012}. 

\begin{figure}[htbp]
  \centering
  \includegraphics[width=0.82\linewidth]{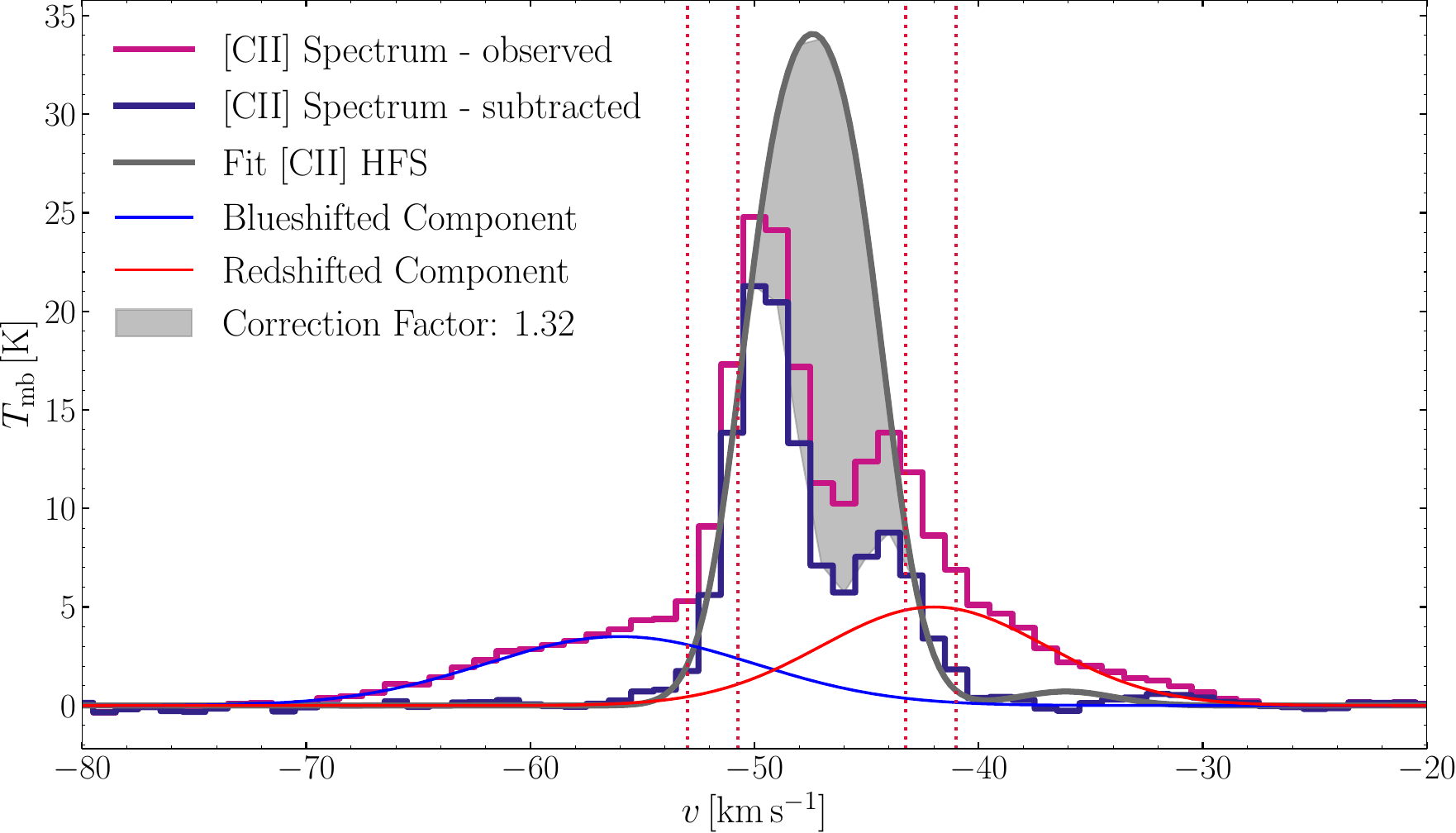}
  \caption
      {Example of the fitting procedure for one spectrum toward the O star with fixed $\Tex=80\,$K. Vertical dotted lines signify the outer wings of the double-peak spectrum.
      The gray area indicates the missing \CII\ emission. The large blueshifted and redshifted expanding shell of RCW79 is fit and indicated. 
      }
    \label{fig:CII_fit_of_wings}
\end{figure}

To correct for the missing emission, we fit the \CII\ line and the \thCII\ HFS to the outer wings of the double-peak spectra using radiative transfer equations, where the optical thickness, peak position, and width were free fit parameters.
We first detected the peaks and then subtracted the blueshifted and redshifted components of the large shell of RCW79 that were fit with Gaussians. For each position, we calculated the area between the spectra and the fit line, denoted as the ``correction factor;'' we show an example spectrum in Fig.~\ref{fig:CII_fit_of_wings}. 
We performed calculations for an excitation temperature, $\Tex$, between 54 to 80$\,$K in steps of $1\,$K, but kept each $\Tex$ fixed during the fitting process. 
As a lower limit, 54$\,$K is derived from the two-layer model (Appendix~\ref{app:two_layer_model}) and 80$\,$K is an upper limit, since the mean correction factor becomes insensitive to $T_\mathrm{ex}\gtrsim80\,$K  
(see Fig.~\ref{fig:mean_correction_factor_vs_Tex}). 
The correction factor is shown in Fig.~\ref{fig:CII_correction_factor_map} and varies from $\sim\,$$1.1$ to $\sim\,$$1.4$.

\begin{figure}[htbp]
  \centering
  \includegraphics[width=0.8\linewidth]{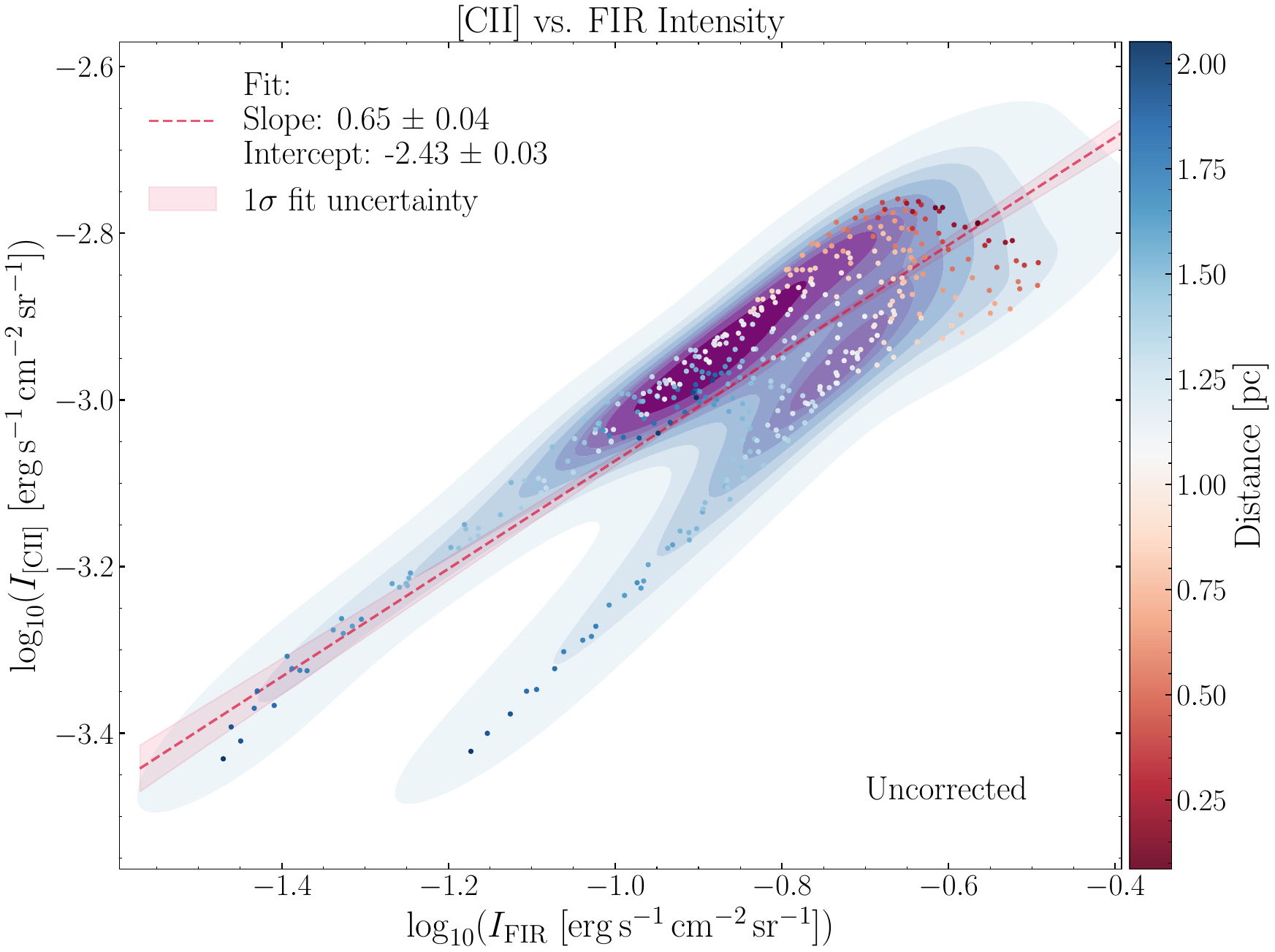}
  \includegraphics[width=0.8\linewidth]{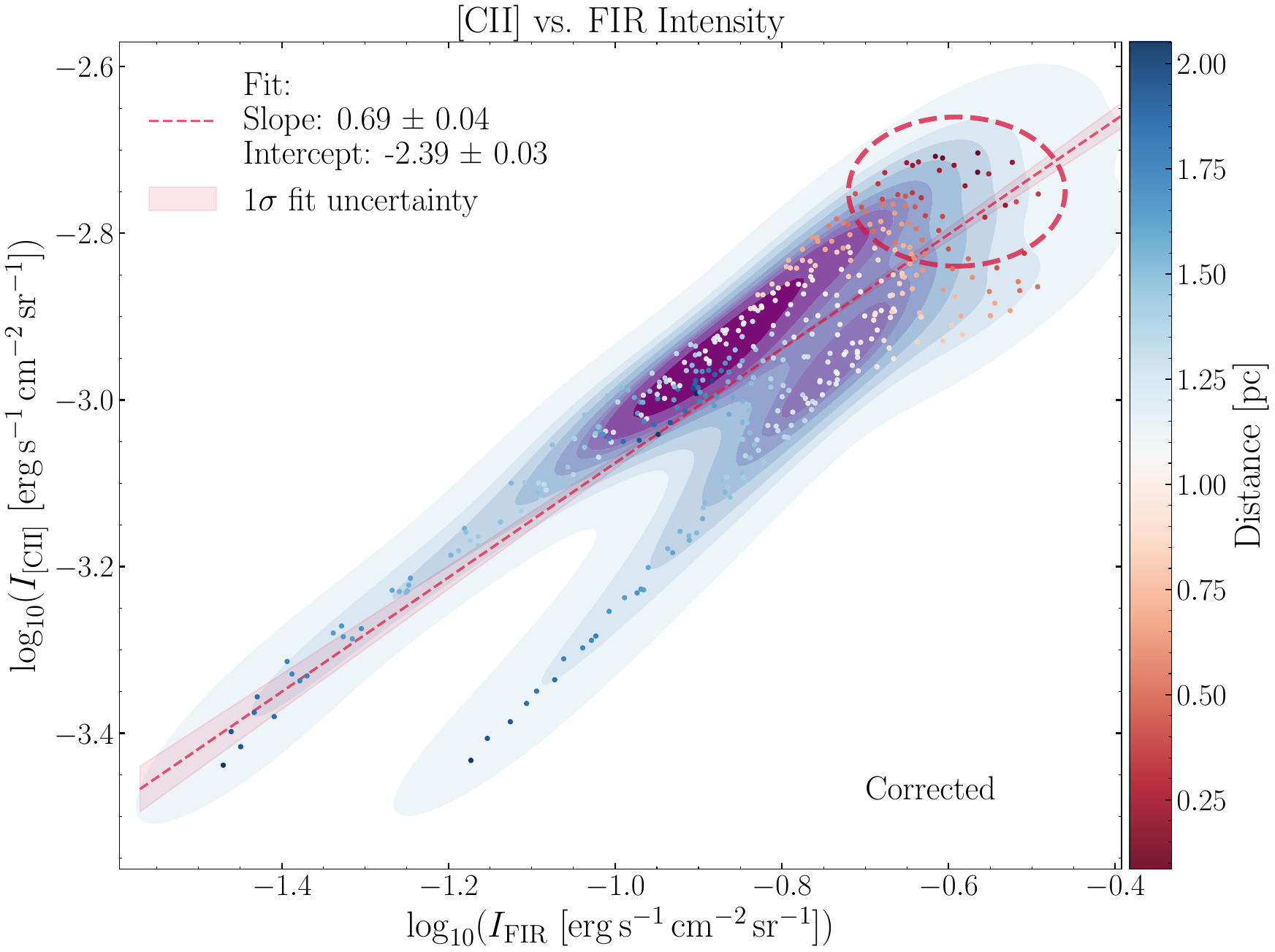}
  \caption 
      {
      \CII/FIR correlation for both uncorrected (top) and corrected (bottom) \CII\ emission. Pixels (derived from maps at 36$''$ resolution) are color-coded based on their proximity to the O star; their density is reflected in the shaded blue areas. A dashed red line represents a linear fit intended to guide the eye for a linear \CII/FIR correlation. 
      Corrected pixels near the O star with high \CII\ and FIR values $\mathrm{log}({\rm FIR})>-0.6$ are shifted upward in the lower panel (indicated by a dashed red ellipse).
      }
    \label{fig:CIIvsFIR}
\end{figure}

Figure~\ref{fig:CIIvsFIR} compares the correlations of the \CII\ intensity with the FIR intensity for both the ``uncorrected'' (top panel) and corrected \CII\ values (bottom panel) reconstructed in the manner described above.\footnote{A linear relation, $\eta=\alpha+\beta\xi+\epsilon~$, in log-log space was fit to the samples using Bayesian Interference with \texttt{linmix}~\citep{Kelly2007}. $\alpha$ and $\beta$ correspond to the intercept and slope, while $\epsilon$ is the intrinsic scatter of the relation. The absolute values, however, are not relevant here and will be discussed in another study.} The total FIR intensity was determined from a spectral energy distribution fit to the \textit{Herschel} fluxes (Appendix~\ref{app:SED_fitting}). 
First, the distribution is bimodal for $\mathrm{log}({\rm FIR})\lesssim-0.9$ and $\mathrm{log}({\rm \CII})\lesssim-3.1$ due to environmental effects. 
The corresponding pixels originate from the map edges in the NW and SE, where PDR gas emits at similar \CII\ levels but exhibits weaker FIR emission in the NW. Consequently, the \CII/FIR ratio is higher in the NW, resulting in the observed bimodality. 

Second, the upper panel shows that the highest \CII\ intensities depend less than linearly on the FIR. This so-called \CII-deficit is sometimes observed in Galactic and extragalactic star-forming regions.
For ultraluminous IR galaxies,
it is often a high dust optical depth, a high ionization parameter (when carbon can become doubly ionized to C$^{2+}$ and higher states), metallicity variations, and very strong and hard radiation fields that can cause the \CII-deficit~(\citealt{Lagache2018},~\citealt{Luhman2003}, and references therein).
In less extreme environments in Galactic PDRs, other processes are proposed. In very dense regions, an increased heating efficiency caused by dust grains 
and collisional de-excitation of \CII\ can take place and in environments with high values of $G_0\, T^{0.5}/n_e$, photoelectric heating efficiency may drop, reducing overall gas cooling~\citep{Hollenbach1999,Goicoechea2015,Pabst2021}. Saturation\footnote{The \CII\ emissivity ceases to increase, even with a stronger radiation field. In this case, the fraction of ions in the upper excited state of the two-level system of the \CII\ 158$\,\mum$ line reaches a threshold.} of the \CII\ line can occur under specific conditions, and \OI\ cooling~\citep{Hollenbach1999} can dominate in very dense and warm PDRs. Self-absorption of the \CII\ line is another viable mechanism for reducing the integrated \CII\ line emission. This has not yet been discussed for \CII\ in the literature; however,~\citet{Goldsmith2021} propose this mechanism for a low \OI/FIR ratio in W3.

For S144, we can exclude processes requiring extremely high density and UV radiation fields, as the conditions are moderate ($n\sim$ a few 1000$\,\mathrm{cm^{-3}}$, $\mathrm{G_0} \sim$ a few 1000).
Saturation is also unlikely, as it typically becomes significant for dust temperatures exceeding $40\,$K (Fig.~16 of~\citealt{Ebagezio2024}).  However, the dust temperatures in S144 are below $30\,$K (Fig.~2a in~\citealt{Liu2017}). 
A high ionization parameter is also not expected to be a significant issue. The \CII\ emission predominantly originates in the PDR, where carbon remains in a C$^+$ state. While the C$^+$ to C$^{2+}$ transition may occur within the \HII\ region,~\citet{Ebagezio2024} indicate that this process is more relevant for evolved \HII\ regions, which is not the case for our \cHIIR. 
\citet{Gerin2015} propose that the \CII-deficit toward nuclear regions in luminous IR galaxies is caused by absorption on kiloparsec scales from diffuse gas in the foreground.
This is not what we suggest, because in our case the absorption takes place very close to the source and is mostly due to a temperature and velocity gradient in the PDR. 

Our straightforward approach of correcting for self-absorption shows that the \CII\ to FIR correlation becomes linear again for high \CII\ and FIR values, as shown in Fig.~\ref{fig:CIIvsFIR}, in which the \CII\ values shift upward.
This finding does not exclude the possibility that other mechanisms are also at work, but demonstrates that in this PDR region with not-so-extreme radiation fields 
and densities  
the \CII-deficit can indeed be mainly caused by self-absorption effects in the \CII\ line. 
This view will be further investigated by studying the \CII/FIR ratio and the \CII\ line properties in other FEEDBACK sources and carrying out a comparison with simulations of \HII\ regions including stellar winds and radiation.

\begin{acknowledgements}
This study was based on observations made with the NASA/DLR  Stratospheric Observatory for Infrared Astronomy (SOFIA). SOFIA is
jointly operated by the Universities Space Research Association Inc. (USRA), under NASA contract NNA17BF53C and the Deutsches SOFIA
Institut (DSI), under DLR contract 50 OK 0901 to the University of Stuttgart. upGREAT is a development by the MPIfR and the KOSMA/University Cologne, in cooperation with the DLR Institut f\"ur Optische Sensorsysteme.
This work was carried out in part at the JPL, California Institute of Technology, under contract with the National Aeronautics and Space Administration. 
Financial support for FEEDBACK at the University of Maryland was provided by NASA through award SOF070077 issued by USRA. The FEEDBACK project was supported by the BMWI via DLR, project number 50OR2217. 
S.D. acknowledges support from the IMPRS
at the Universities of Bonn and Cologne.
S.K. acknowledges support by the BMWI via DLR, project number 50OR2311. 
This work was supported by the CRC 1601 (SFB 1601 sub-projects B2, C1, C2, C3, C6) funded by the DFG – 500700252.
\end{acknowledgements}  

\bibliographystyle{aa}
\bibliography{aandaRCW79}

\begin{appendix}
\section{Multiwavelength plots and channel maps in \CII\ and $^{12}$CO 3$\,\to\,$2 emission}
\label{app:complementary_data}

\begin{figure*}[!h]
  \centering
  \includegraphics[width=0.41\linewidth]{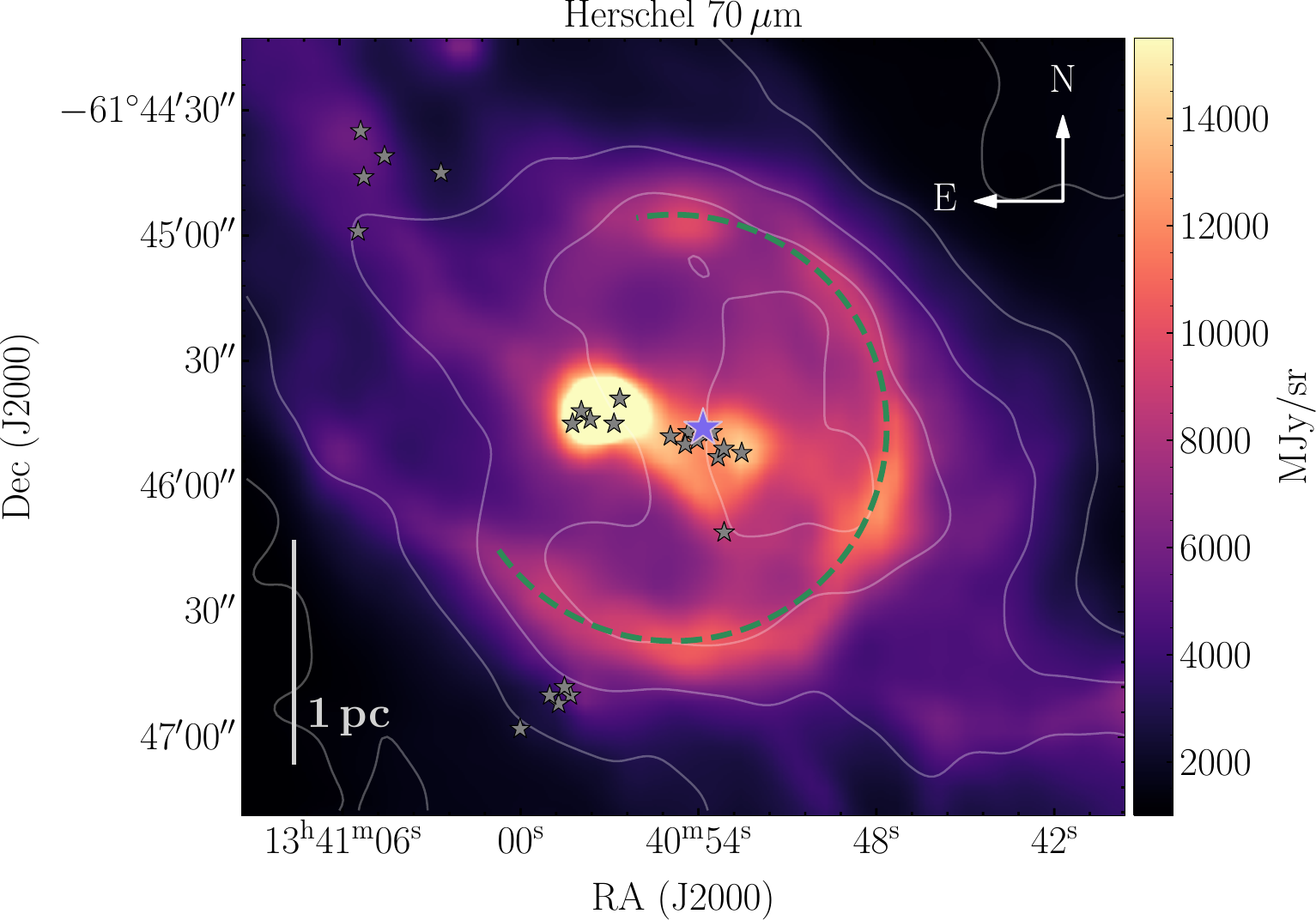}
  \includegraphics[width=0.39\linewidth]{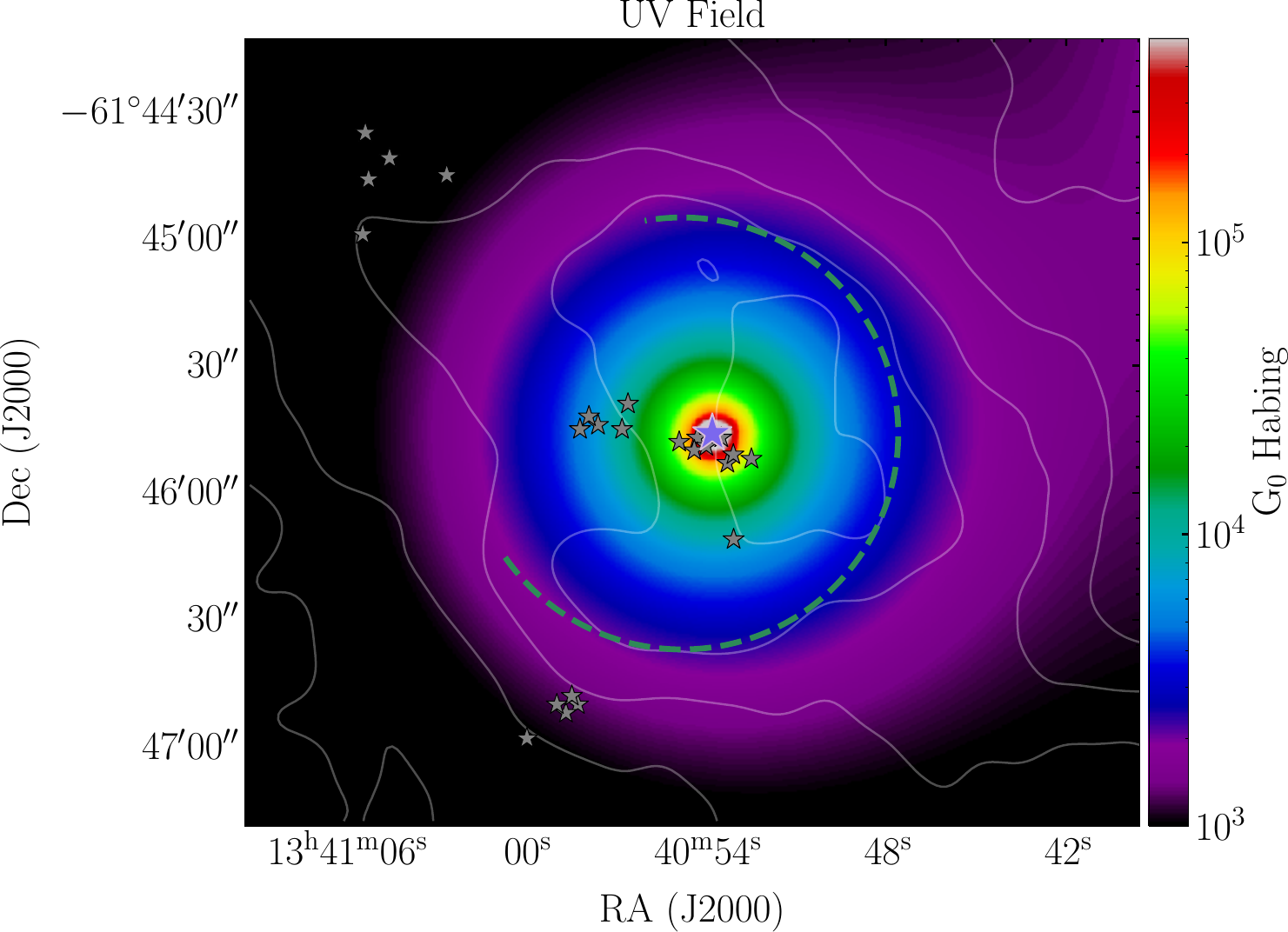}
  \includegraphics[width=0.41\linewidth]{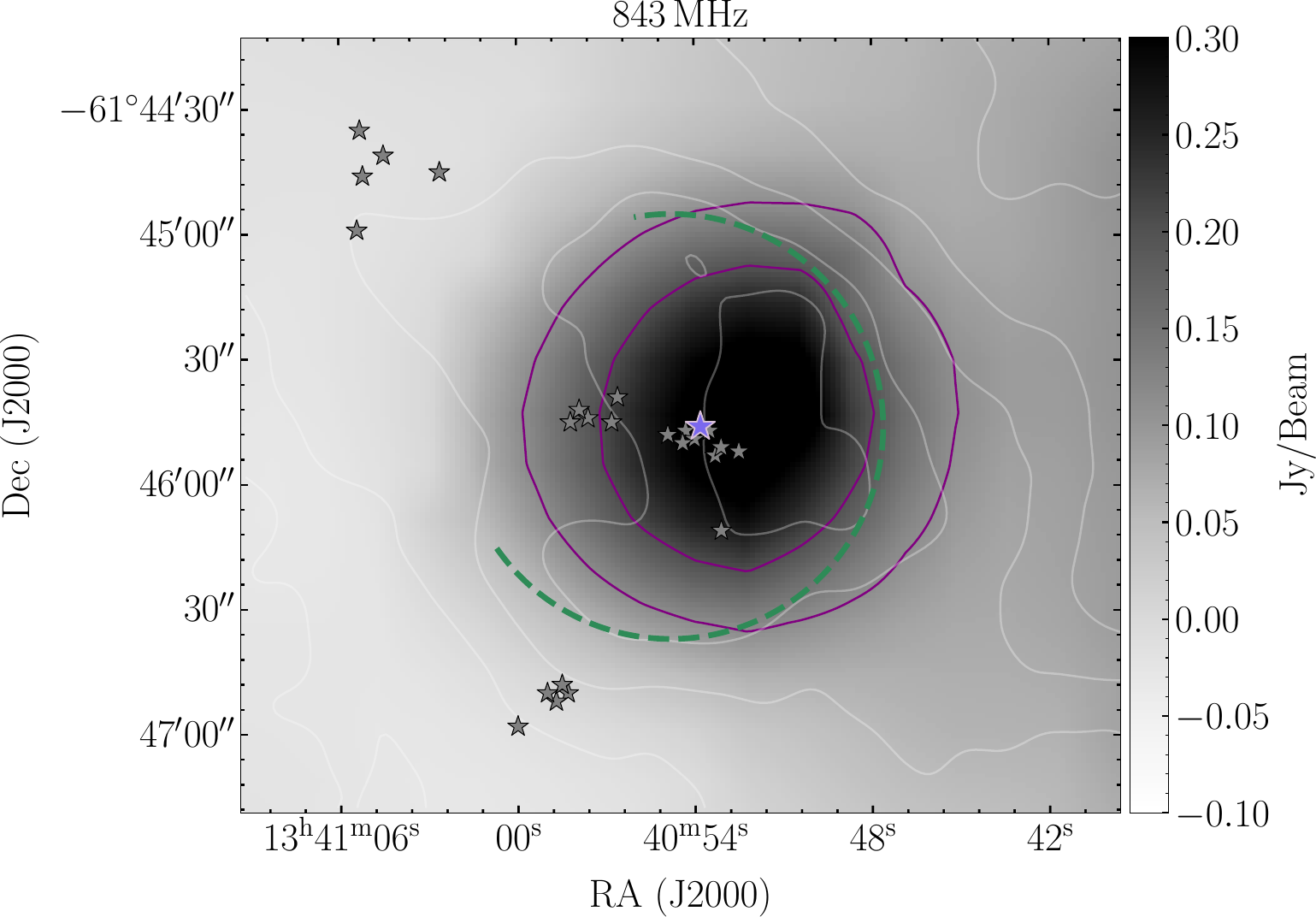}
  \includegraphics[width=0.39\linewidth]{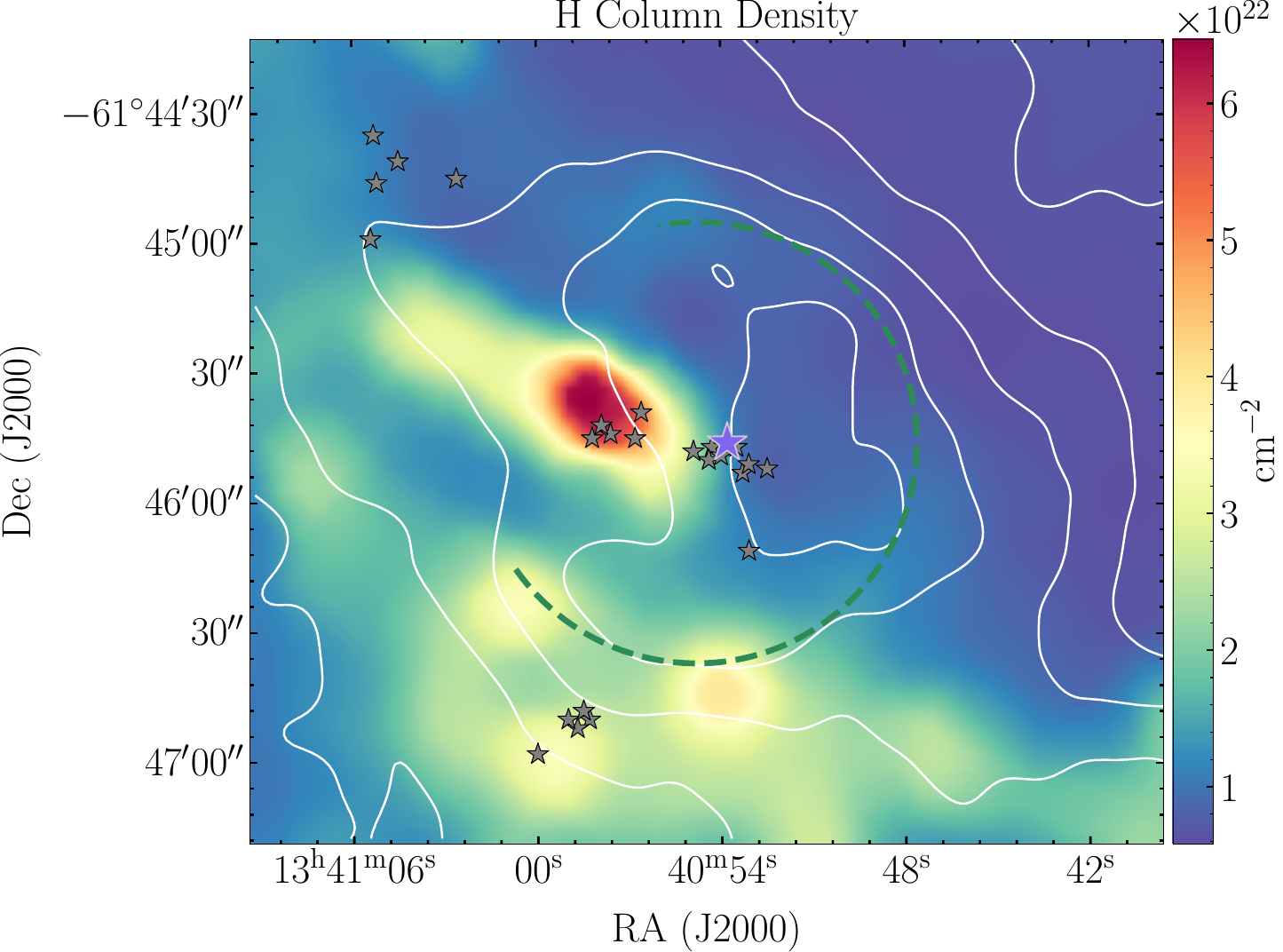}
  \caption
      {Complementary plots for the S144 C$^+$ bubble in RCW79. 
       The upper left panel illustrates the $70\,\mum$ emission from \textit{Herschel} at 
       $\sim\,$$6''$ resolution. This emission resembles the \textit{Spitzer} $8\,\mum$ image (as seen in Fig.~\ref{fig:spitzer}), though it highlights more clearly the NE clump containing the embedded IR cluster in the $70\,\mum$ map. The upper right panel presents the UV field, which is derived from the spectral type of the O star (indicated by a blue star) and the central O-cluster from RCW79, located farther to the west. The lower left panel displays the 843$\,\mathrm{MHz}$ emission with a $45''$ angular resolution, outlining the \HII\ region~\citep{Cohen2002}. In the lower right panel, the \textit{Herschel} dust column density map~\citep{Liu2017} at 18$''$ angular resolution reveals the clumpy structure of condensation 2. Each panel includes \CII\ contours in increments of $50\,\kkms$ from 50 to 250$\,\kkms$, while the 843$\,\mathrm{MHz}$ map also depicts 843$\,\mathrm{MHz}$ contours at $0.1$ and $0.2\,$Jy/Beam in purple. The small black stars are members of the IR clusters.
      }
    \label{fig:complementary_plots}
\end{figure*}

\begin{figure*}[!h]
  \centering
  \includegraphics[width=0.81\linewidth]{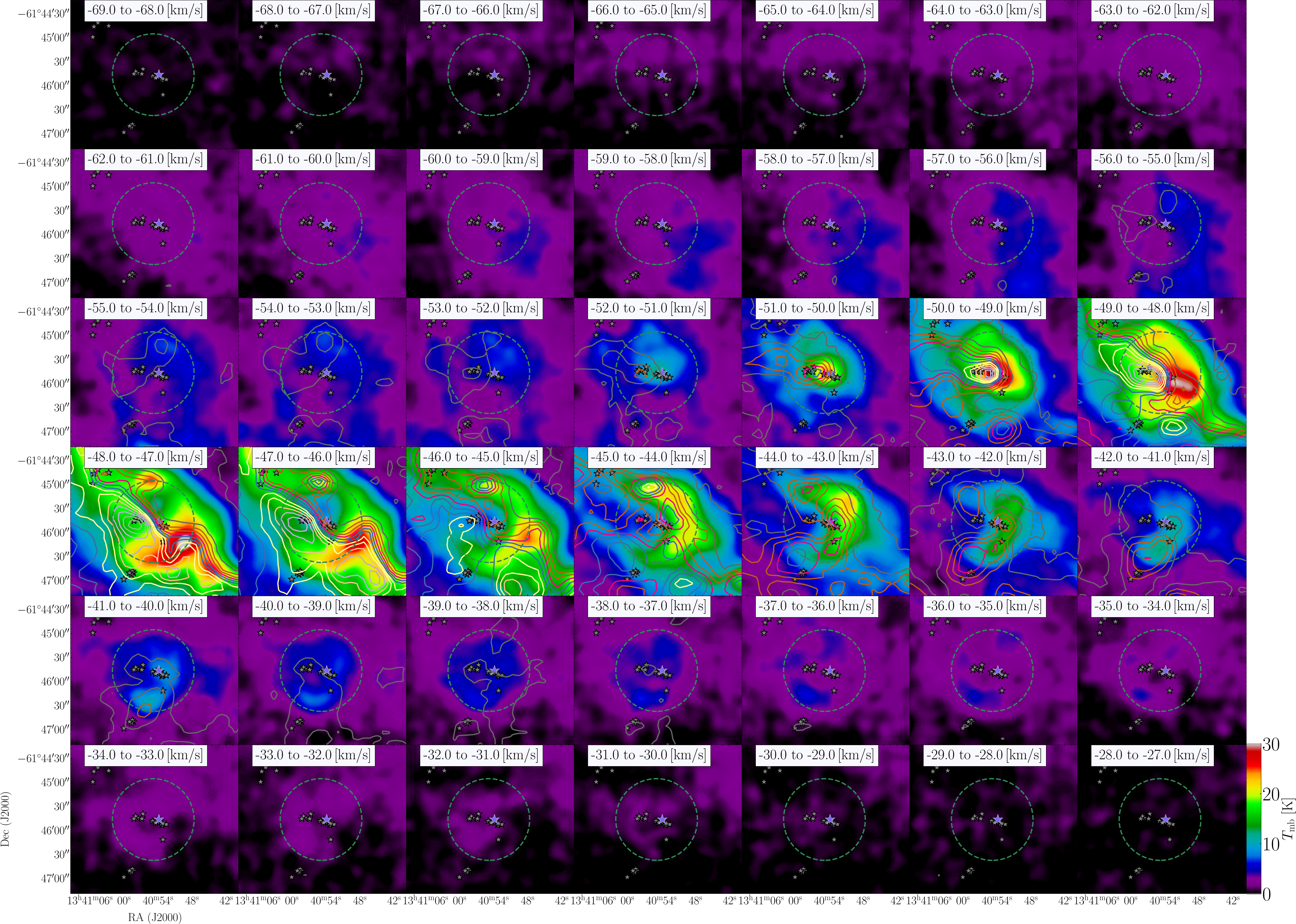}
  \caption
      {Channel map showing \CII\ and $^{12}$CO 3$\,\to\,$2 emissions of the C$^+$ bubble. CO contours range from 0.67 to 32$\,\kkms$ in steps of 2.4$\,\kkms$. The prominent blue star marks the position of the exciting O star, while smaller black stars denote IR cluster members. The green dotted circle indicates the approximate extent of the IR shell as observed in the \textit{Spitzer} $8\,\mum$ image. 
      }
    \label{fig:CII_channel_map_co32Contours}
\end{figure*}
\begin{figure*}[!h]
  \centering
    \includegraphics[width=0.45\linewidth]{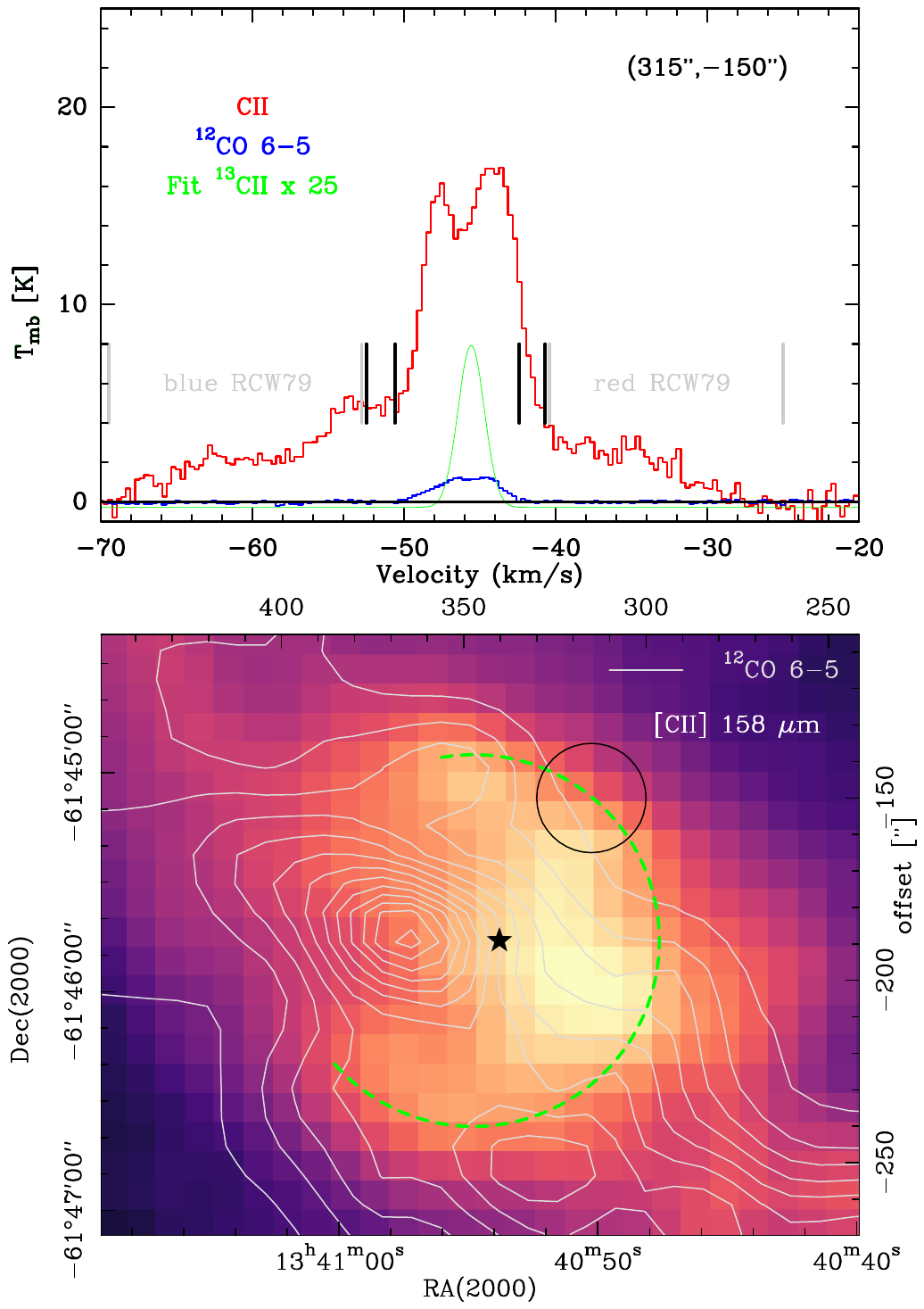}
    \includegraphics[width=0.45\linewidth]{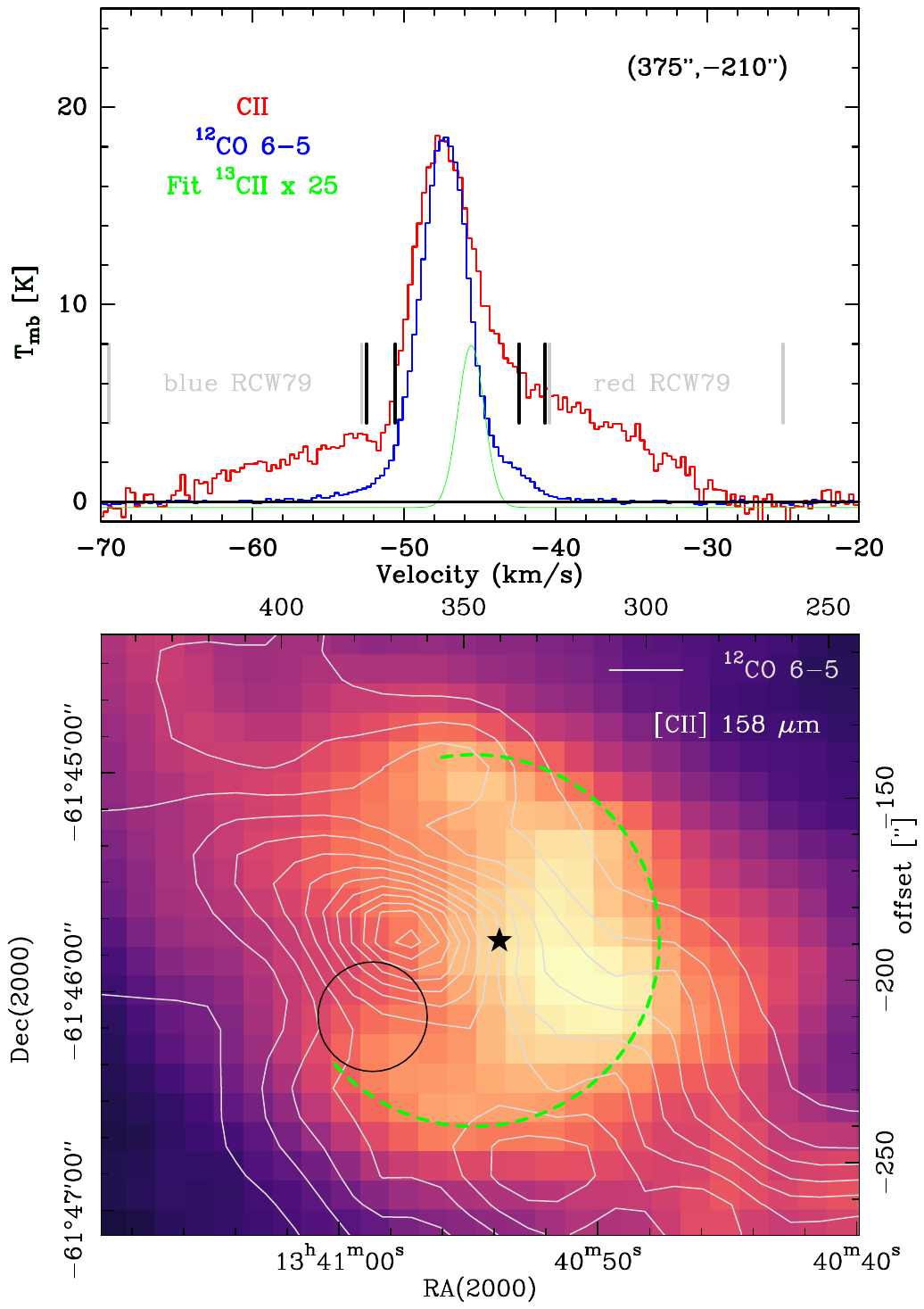}
  \caption
      {
      Example spectra of \twCII\ and $^{12}$CO\,6$\,\to\,$5 lines with a fit to the HFS $\mathrm{F}=1-0$ component of the average \thCII\ line. 
      The upper panels show the spectra/fit in a smoothed 30$''$ beam at the positions marked by a black circle in the lower panel of line-integrated \CII\ emission and $^{12}$CO\,6$\,\to\,$5 emission contours (10 to 130 by 10$\,\kkms$). 
      The two pairs of gray lines mark the blue and red velocity ranges of the large expanding \CII\ shell over the whole RCW79 region~\citep{Bonne2023}. The two pairs of black lines outline approximately the velocity range of the expanding \CII\ shell. 
      The black star in the lower panels indicates the position of the O-star and the green dashed circle the 8$\,\mum$ ring. 
      A movie showing all spectra is available \href{https://hera.ph1.uni-koeln.de/~nschneid/feedback.html}{online}. 
      }
    \label{fig:spectra}
\end{figure*}

Figure~\ref{fig:complementary_plots} presents a \textit{Herschel} $70\,\mum$ map, our computed UV field map, and a $843\,\mathrm{MHz}$ emission and \textit{Herschel} column density map. The $843\,\mathrm{MHz}$ map outlines  the approximate extent of the \HII\ region, though it suffers from beam dilution due to its 45$''$ beam size. The dust ring, seen in both the \textit{Herschel} $70\,\mum$ and the \textit{Spitzer} $8\,\mum$ maps (Fig.~\ref{fig:spitzer}), probably indicates compressed gas, predominantly to the west. 
The FUV map of the total RCW79 region was generated following the procedure outlined in~\citet{Schneider2023}, using the spectral classification of all O stars as detailed in~\citet{Martins2010}. 
The features in these maps partly motivated the modeling of the two outer shells using SimLine (Sect.~\ref{app:SimLine}).

The \textit{Herschel} dust column density map~\citep{Liu2017} illustrates the dense gas distribution, highlighting a SE region comprising multiple high-density clumps, one reaching up to $6\times10^{22}\,\mathrm{cm^{-2}}$ NE of the O star. Identified as a cluster-forming star-formation site~\citep{Zavagno2006}, this molecular clump shields the \HII\ region and PDR gas. The movie overlays with \CII\ and CO alongside the column density nicely show the gas flow around this clump.

Figure~\ref{fig:CII_channel_map_co32Contours} displays a \CII\ channel map with superimposed contours of $^{12}$CO 3$\,\to\,$2 emission. The purple areas represent \CII\ emission of low surface brightness, primarily originating from the large expanding \CII\ shells spanning the RCW79 region and the disorganized \CII\ flows that are eroding the molecular cloud~\citep{Bonne2023}. Within the velocity range of $-54\,\kms$ to $-40\,\kms$ (indicated in blue), we detect the shell/bubble of \CII\ exclusively associated with the \cHIIR. The ``filled C$^+$ bubble'' geometry is disturbed by the prominent CO clump, which becomes visible around $-59\,\kms$ to the west. Furthermore, near a velocity of $\sim\,$$-47\,\kms$, a \CII-deficit is detected due to self-absorption.

Figure~\ref{fig:spectra} displays a map of the line-integrated \CII\ and $^{12}$CO\,6$\,\to\,$5 emission, together with spectra overlays selected for two representative positions. The NW spectrum (left) is an example of a self-absorbed \CII\ line, also showing prominent wings from the expanding \CII\ shell. The right panel is a position from the SE where the PDR emission from the surface of the molecular clump dominates the \CII\ and CO emission. The high-velocity \CII\ wings are, however, well visible.

\section{The two-layer radiative transfer model} \label{app:two_layer_model}

\begin{figure*}[htbp]
  \centering
  \includegraphics[width=0.75\linewidth]{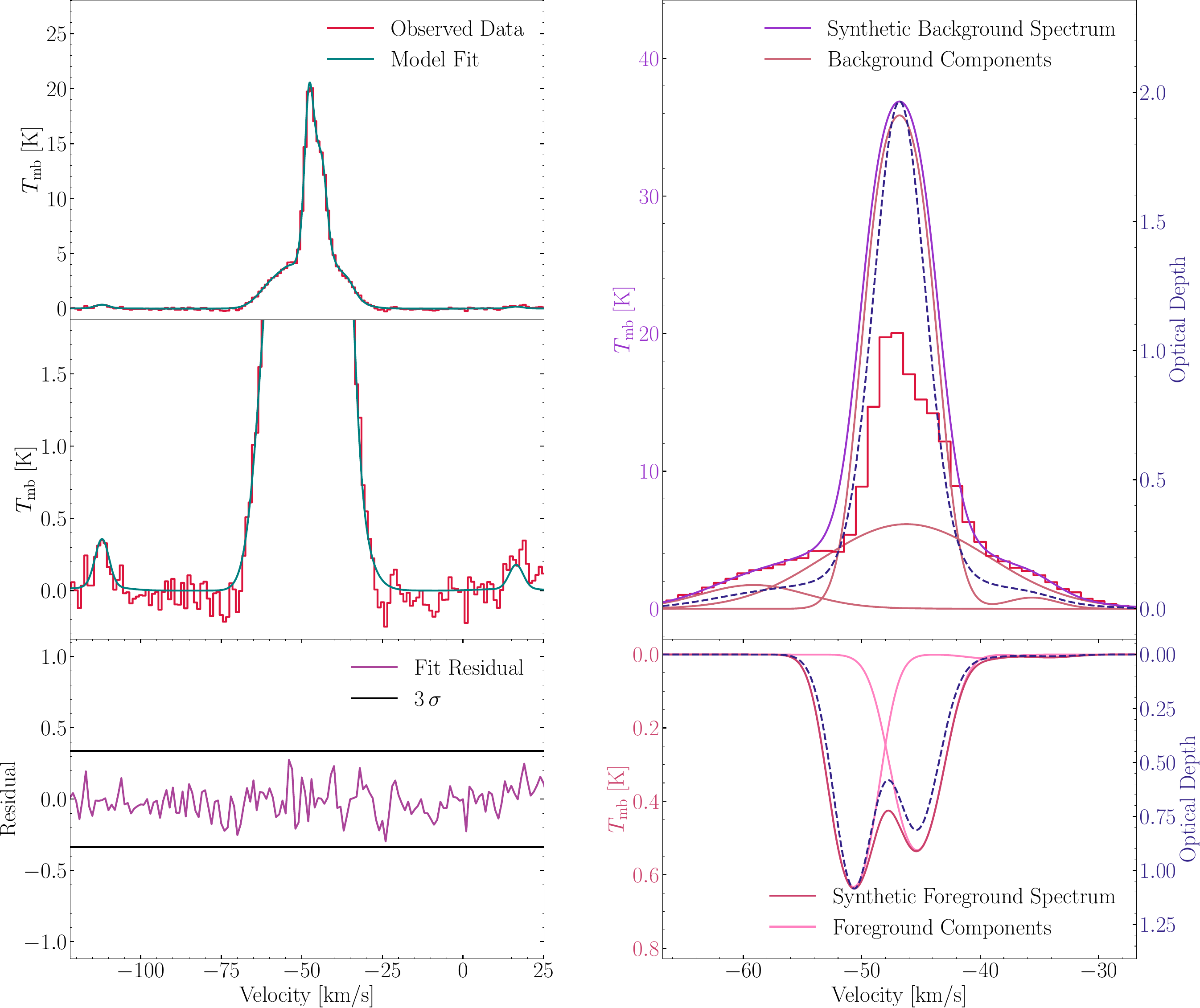}
  \caption
      {Two-layer model results for the average \CII\ spectrum. 
      The average spectrum corresponds to the area defined with dendrograms (refer to Fig.~\ref{fig:cii_integrated_map}, right). The spectrum, shown in red, has a velocity resolution of $1\,\mathrm{km\,s^{-1}}$. The model's background and foreground layer temperatures are $T_{\rm ex, bg}=80\,$K and $T_{\rm ex, fg}=20\,$K respectively. The upper left panel illustrates the observed spectrum in red beside its two-layer model fit in blue. The left central panel replicates the top panel's data but expands the intensity axis for clearer viewing of the three \thCII\ lines. The bottom left panel shows the residuals of the fit, with two horizontal black lines denoting the $3\sigma$ threshold. The top right panel focuses on individual background elements in red and the total background in violet. Meanwhile, the dashed blue curve represents velocity-resolved optical depth. The bottom right panel illustrates individual foreground elements in pink, alongside the overall foreground in red, again accompanying the dashed blue curve for optical depth representation.
      }
    \label{fig:CII_two_layer_model}
\end{figure*}

The two-layer multicomponent model~\citep{Guevara2020,Kabanovic2022} solves the radiative transfer equation for multiple velocity components distributed between two layers with different excitation temperatures along the line-of-sight. While the model itself has no restriction on the excitation temperature, we choose the excitation temperature $T_{\rm{ex, bg}}$ of the background layer (bg) to be higher than the temperature $T_{\rm{ex, fg}}$ of the foreground layer (fg), considering the geometry of the region with the exciting O star in the center and cooler PDR shells around. Thus, we account for self-absorption due to a column of gas along the line-of-sight in a single layer (which results in flat top spectra for high optical depth $\tau$) as well as the foreground absorption by a colder foreground layer (which produces absorption dips in the spectrum). The absorption can originate from a spatially separated, cold foreground cloud, which is located between the warm emitting gas and the observer, or from a temperature gradient along the line-of-sight of the same cloud.
Following the model presented in~\cite{Kabanovic2022}, we solve the following equation:
\begin{equation}  \label{eq:2layer} 
    \begin{split}
        T_{\mathrm{mb}}(v) = & \left[ \mathcal{J}_{\nu}(T_{\mathrm{ex},\mathrm{bg}}) \, \left( 1-\e^{-\sum_{i_\mathrm{bg}}\tau_{i_\mathrm{bg}}(v)}\right) \right] \e^{-\sum_{i_\mathrm{fg}} \tau_{i_\mathrm{fg}}(v)} + \\
        &  \mathcal{J}_{\nu}(T_{\mathrm{ex},\mathrm{fg}}) \, \left( 1-\e^{-\sum_{i_\mathrm{fg}}\tau_{i_\mathrm{fg}}(v)}\right)~. 
    \end{split} 
\end{equation} 
\noindent
The equivalent brightness temperature of a black body emission at a temperature $T_{\mathrm{ex}}$ is 
\begin{equation}
    \mathcal{J}_{\nu}(T_{\mathrm{ex},i}) = \frac{T_0}{\e^{T_0/T_{\mathrm{ex}}}-1}~,
\end{equation}
\noindent
with the equivalent temperature of the transition $T_0=h\nu/k_B$ and $\nu$ the transition frequency. The optical depth of each Gaussian component is given by
\begin{equation}
    \tau(v)=\tau_0 \,\e^{-4\ln 2~\left(\frac{v-v_{0}}{w}\right)^2}~,
\end{equation}
\noindent
with the central local standard of rest (LSR) velocity $v_{0}$ of each component. The line width $w$ is expressed as the full width at half maximum (FWHM) of the component. For a simple two-level system, we can express the peak optical depth $\tau_0$ of each Gaussian component as a function of the excitation temperature $\Tex$ and column density $N$ by 
\begin{equation}
    \tau_0 = N\frac{c^3}{8\pi\nu^3}\frac{g_u}{g_l}A_{ul}\frac{1-\e^{-T_0/\Tex}}{1+\frac{g_u}{g_l}\e^{-T_0/\Tex}}\frac{2\sqrt{2\ln2}}{w\sqrt{2\pi}}~.
    \label{eq:column_density_two}
\end{equation}
\noindent
For the \CII\ fine-structure transition, the rest frequency is $\nu = 1900.5369\,\mathrm{GHz}$, the Einstein coefficient for spontaneous emission  $A_{ul} = 2.29\cdot10^{-6}\,\mathrm{s^{-1}}$,  the equivalent temperature of the upper level $T_0=h\nu/k_{\rm B}=91.25\,\mathrm{K}$, and the statistical weights of the transition energy levels are $g_u=4$ and $g_l=2$. 
\begin{table}[htb!]
            \caption{Two-layer multicomponent model results.}
            \label{tab:two-layer}
            \begin{tabular}{l|rrrr}
                \hline 
                \hline
                & \multicolumn{4}{|c}{\bf Model I: $T_{\rm ex, bg} = 54\,{\rm K}$, $T_{\rm ex, fg} = 20\,{\rm K}$} \\
                \hline
                Components & $N_{\rm \CII}$  &$\tau_{\mathrm{0}}$  & $v_0$ & $w$ \\
                & [$10^{18}\,{\rm cm^{-2}}$] &  & [km$\,$s$^{-1}$] & [km$\,$s$^{-1}$]\\
                \hline
                Backg. Comp. 1 & $4.77$ & $4.26$ & $-47.06$ & $4.48$\\
                Backg. Comp. 2 & $1.37$ & $0.27$ & $-48.88$ & $20.02$\\
                Backg. Comp. 3 & $0.54$ & $0.32$ & $-44.29$ & $6.67$\\
                \hline
                Foreg. Comp. 1 & $0.17$ & $0.28$ & $-44.96$ & $2.36$\\
                Foreg. Comp. 2 & $0.64$ & $0.89$ & $-50.88$ & $2.90$\\
                \hline
                & \multicolumn{4}{|c}{\bf Model II: $T_{\rm ex, bg} = 80\,{\rm K}$, $T_{\rm ex, fg} = 20\,{\rm K}$} \\
                \hline
                Backg. Comp. 1 & $3.29$ & $1.81$ & $-46.81$ & $5.07$\\
                Backg. Comp. 2 & $0.16$ & $0.04$ & $-59.13$ & $10.50$\\
                Backg. Comp. 3 & $0.88$ & $0.15$ & $-46.26$ & $15.95$\\
                \hline
                Foreg. Comp. 1 & $1.28$ & $0.81$ & $-45.33$ & $4.43$\\
                Foreg. Comp. 2 & $1.52$ & $1.07$ & $-50.69$ & $3.95$\\
                \hline

                \hline
            \end{tabular}
\end{table}

\begin{figure}[htbp]
  \centering
  \includegraphics[width=0.7\linewidth]{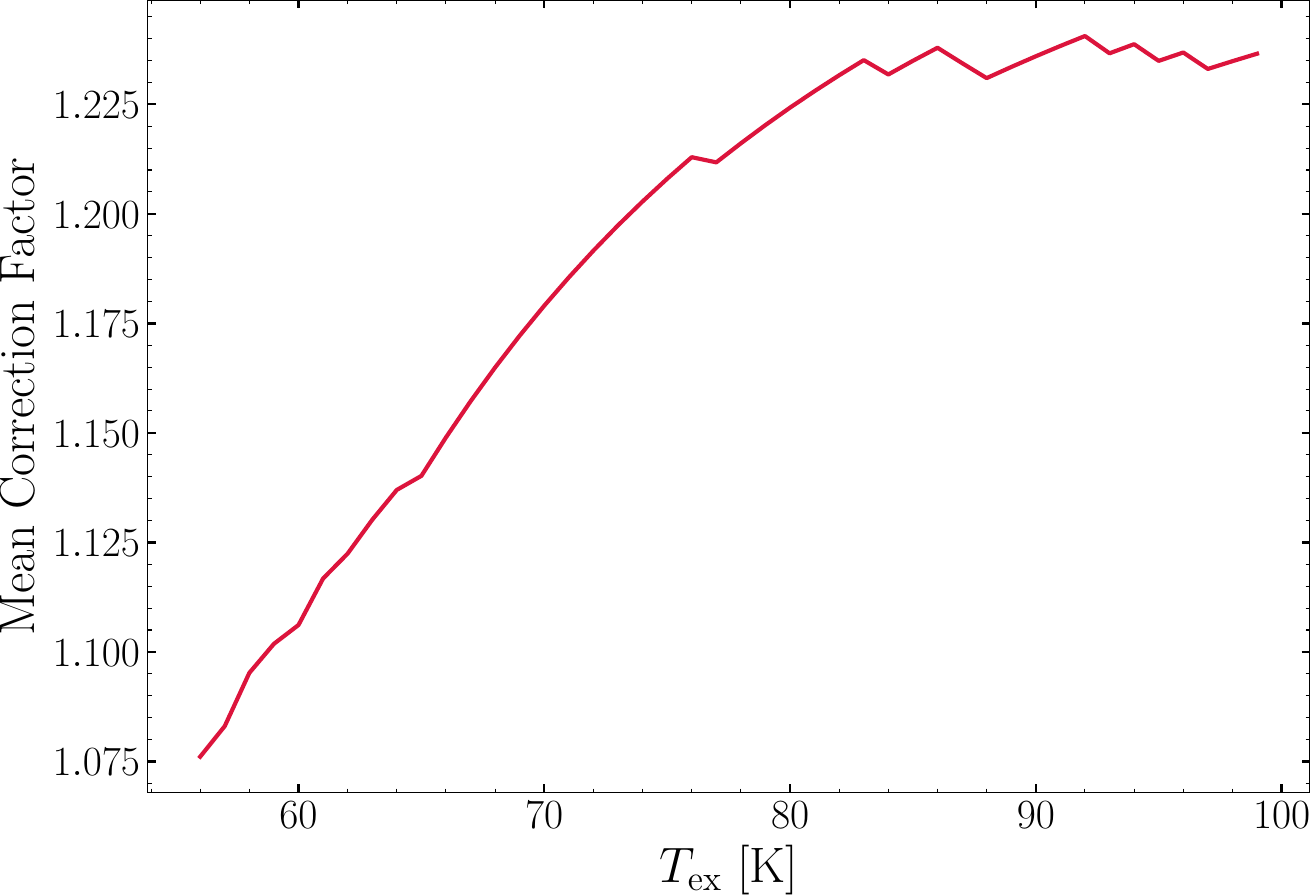}
  \caption
      {Mean correction factor versus $\Tex$. The mean correction factor rises up to $\sim\,$$80\,$K, beyond which it becomes independent of $\Tex$.}
    \label{fig:mean_correction_factor_vs_Tex}
\end{figure}

The physical properties of the background layer can be derived from an optically thin line, which is not affected by self-absorption. In case of \CII, we can utilize the much weaker \thCII\ hyperfine transition lines, see~\cite{Guevara2020} for a more detailed description. Model fit parameters such as LSR velocity, line width, and number of components can be simply derived from the observed line. However, the optical depth and the excitation temperature are not independent of each other. We therefore need to first derive the excitation temperature from the observed data, which leaves the optical depth as the free model fit parameter. Assuming that the warm emitting background can partly shine through the cold absorbing layer~\citep{Kabanovic2022}, we can determine the excitation temperature via
\begin{equation}
    T_{\mathrm{ex}} = \frac{T_0}{\ln\left(\frac{T_0}{T_{\mathrm{\CII, peak}}}\left(1-\e^{-\tau_{\rm \CII, peak}}\right)+1\right)}~.
    \label{eq:excitation_temperatur}
\end{equation}
\noindent
The optical depth is then calculated from the observed \twCII/\thCII\ ratio at the peak of the \twCII\ emission
\begin{equation}
   \frac{T_{\rm [^{\rm 12}C\,{\tiny II}]}(v)}{T_{\rm [^{\rm 13}C\,{\tiny II}]}(v)} = \frac{1 - \e^{-\tau(v)}}{\tau(v)/\alpha} = \frac{1-\e^{\tau(v)}}{\tau(v)}\alpha~,
   \label{eq:tau_cii}
\end{equation}
\noindent
with the local carbon abundance ratio $\alpha$ that we take as $59\pm10$~\citep{Milam2005}. The resulting velocity-resolved optical depth is shown in Fig.~\ref{fig:CII_average_spec_opticalDepth}. For the calculation, only the second strongest \thCII\ $\mathrm{F}=1-0$ line is used, since the strongest $\mathrm{F}=2-1$ HFS line is covered by the redshifted \CII\ wing and the weakest HFS $\mathrm{F}=1-1$ line is detected only marginally. The resulting lower limit for the \CII\ excitation temperature is $T_{\rm ex} = 54\,{\rm K}$. For the upper limit, we derive an excitation temperature of $T_{\rm ex} = 80\,{\rm K}$, see Fig.~\ref{fig:mean_correction_factor_vs_Tex}. 

\begin{figure}[htbp]
  \centering
  \includegraphics[width=0.75\linewidth]{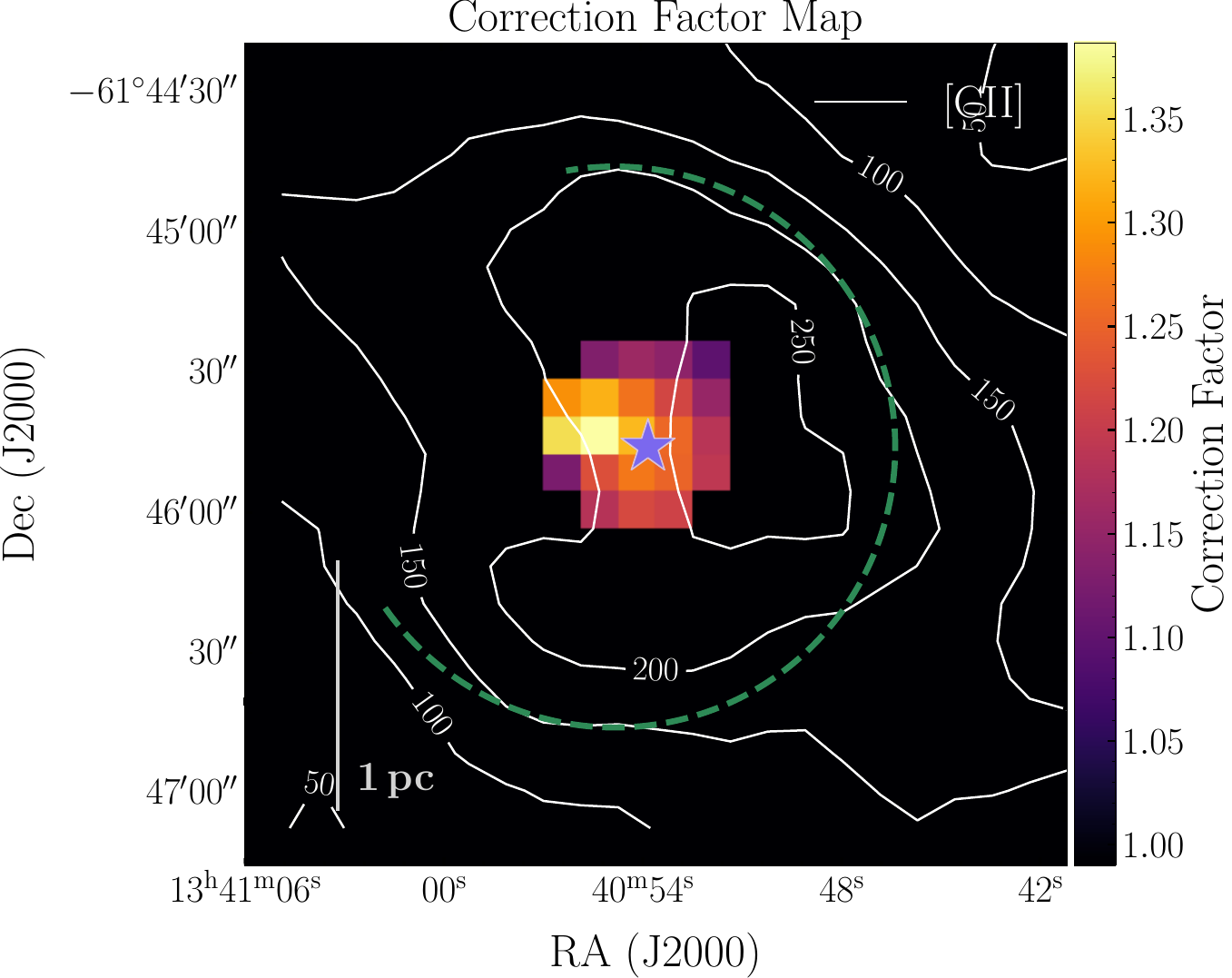}
  \caption
      {\CII\ correction factor map (at an angular resolution of 36$''$ and 8$''$ grid) with fixed $\Tex=80\,$K. 
      }
    \label{fig:CII_correction_factor_map}
\end{figure}
The resulting model fit of the two-layer model is shown in Fig.~\ref{fig:CII_two_layer_model} for a background excitation temperature of $T_{\rm ex, bg} = 80\,{\rm K}$, which is the derived upper limit, indicating an 
optically thick \CII\ line. However, the background (see Table~\ref{tab:two-layer}) derived from the \thCII\ line still overshoots the observed line, which requires additional cold foreground material. Although we do not have observational constraints on the excitation temperature in the foreground, multiple studies~\citep{Kabanovic2022, Schneider2023} have shown that a temperature of $T_{\rm ex, bg} = 20\,{\rm K}$ is reasonable for \CII. We find that the observed foreground material is either blueshifted or redshifted from the systemic velocity. The blueshifted component can be explained due to the expanding bubble, which pushes the cold material in front of it toward the observer. However, the redshifted component cannot be attributed to the opposite hemisphere, since it is only visible in absorption. 
Decreasing the background excitation temperature results in a higher background column density and therefore higher optical depth; see Table~\ref{tab:two-layer}. Note, however, that we derived a temperature of $\sim\,$$100\,$K from the SimLine modeling, so that low excitation temperatures are unlikely. 

As a final note, we show in Fig.~\ref{fig:CII_correction_factor_map} a map of the correction factor that was applied to the spectra showing self-absorption effects. Obviously, the spectra with the highest \CII\ brightness are the most affected.


\section{SimLine modeling of a C$^+$ filled bubble}
\label{app:SimLine}

\begin{figure*}[htbp]
 \centering
 \includegraphics[width=0.33\linewidth]{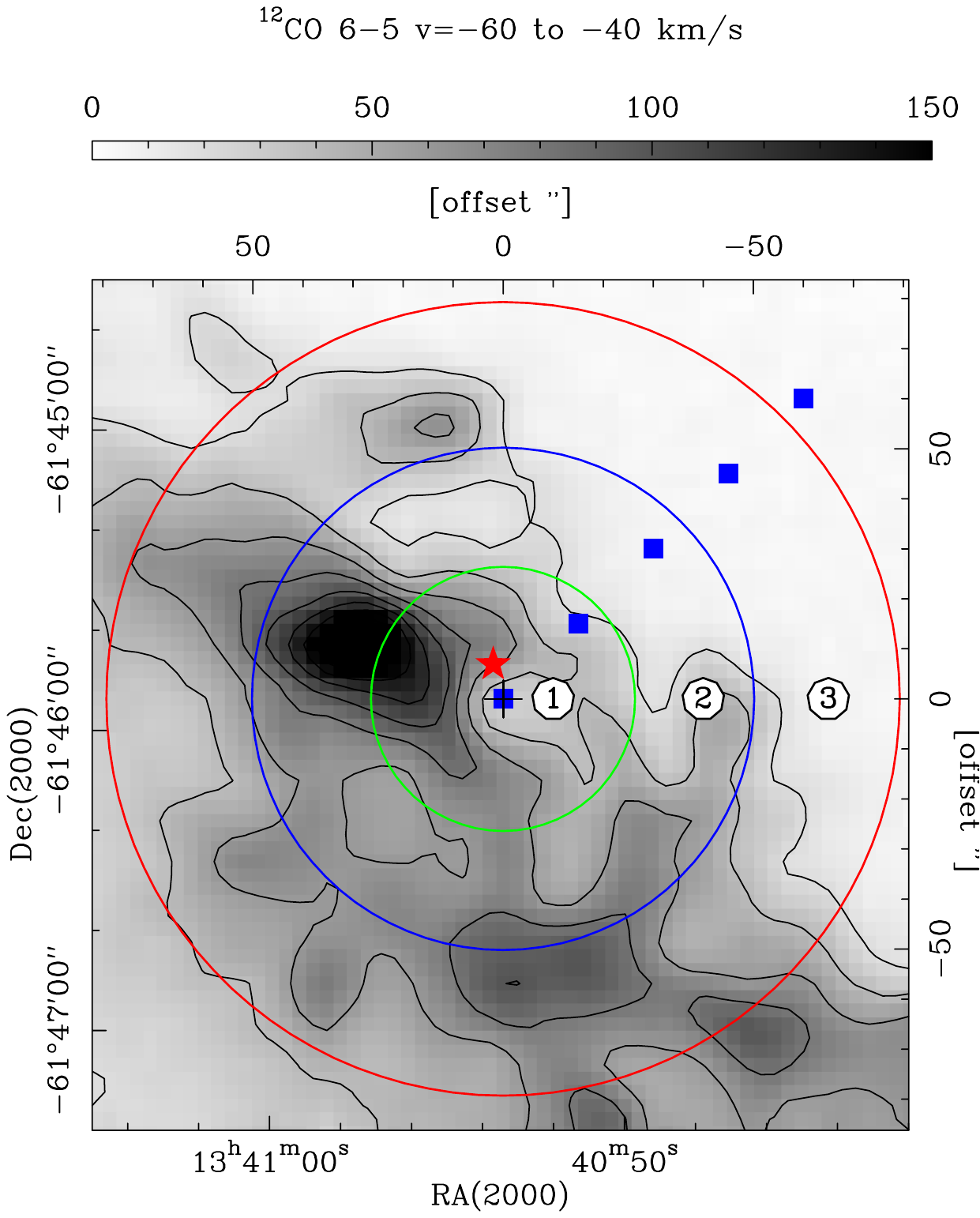}
 \includegraphics[width=0.33\linewidth]{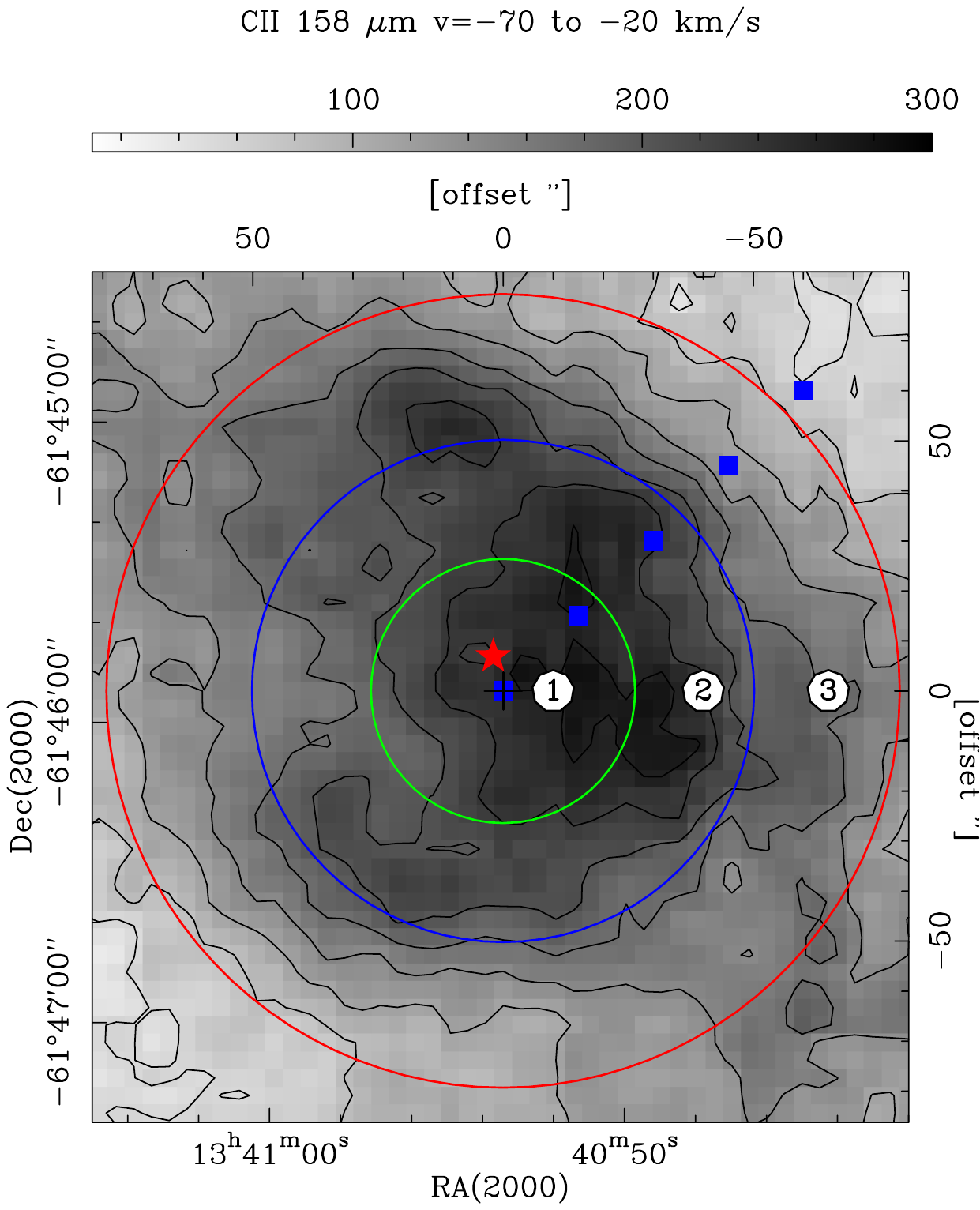}
 \caption
      {SimLine modeling setup of the C$^+$ bubble overlaid on the observed $^{12}$CO 6$\,\to\,$5 and \CII\ intensities. The panels ($^{12}$CO 6$\,\to\,$5 on the left and \CII\ on the right) show a sketch of the setup for an ideal 3D bubble (note that the calculations are in 1D) in which the different shell regions are indicated. The positions of observed and modeled spectra, shown in Fig.~\ref{fig:SimLine2}, are indicated by blue squares. The red star marks the exciting O star of the \cHIIR. }
    \label{fig:SimLine1}
\end{figure*}

\begin{table}[htbp] 
\caption{Parameters of SimLine modeling.} 
\begin{tabular}{lccc} 
\hline 
\hline 
                                                   & shell 1   & shell 2 & shell 3 \\ 
\hline 
{\small outer radius [pc]}                         & 0.5       & 0.95    & 1.5  \\ 
{\small density [cm$^{-3}$]}                       & 100       & 2500    & 2500 \\  
{\small density profile ($\alpha_{s}$)}            & -         &  -      & $-$2  \\ 
{\small temperature [K]}                           & 8000      &  100    & 100  \\  
{\small temperature profile ($\alpha_{s}$)}        &  -        &  -      & $-$2 \\  
{\small turbulent velocity [$\kms$]}               & 3.7       & 2.2     & 2.2 \\ 
{\small turbulent velocity profile ($\alpha_{s}$)} & -         & -       & $-$1 \\ 
{\small radial velocity [$\kms$]}                  & -         & 2.6       & 2.6 \\ 
{\small radial velocity profile ($\alpha_{s}$)}    & -         & -       & $-$2 \\ 
{\small X(\CII) [$10^{-4}$]}                       & 1.2       & 1.2     & 1.2 \\ 
\end{tabular} 
\label{tab:SimLine} 
\end{table} 

\begin{figure*}[htbp]
 \centering
 \includegraphics[width=0.44\linewidth]{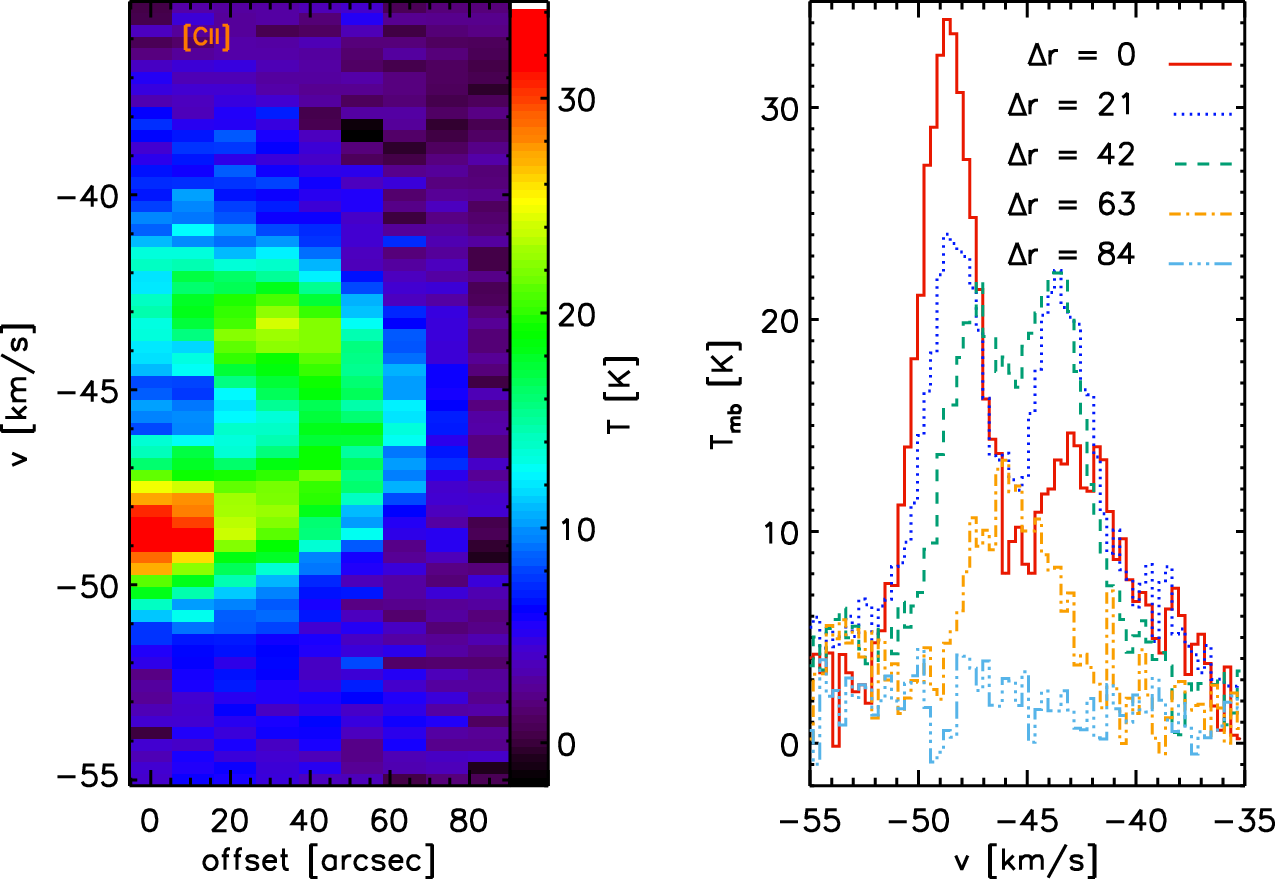}
 \includegraphics[width=0.44\linewidth]{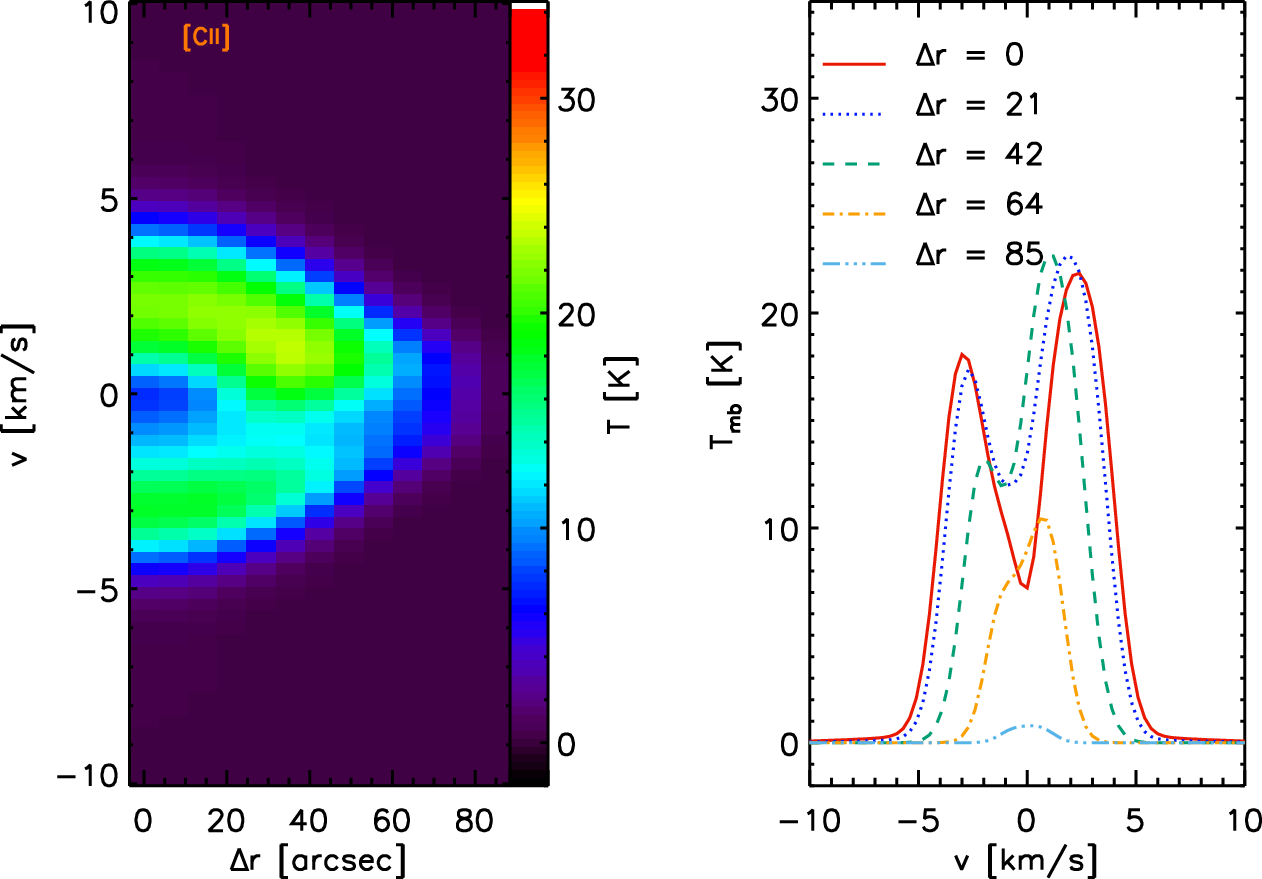}
 \caption
      {Observations versus results of the SimLine modeling. The left panel shows a PV cut from the center position along the positions of the observed and modeled spectra, indicated in Fig.~\ref{fig:SimLine1}, together with the observed spectra. $\Delta r=21$, etc. corresponds to an offset of ($-15''$,$15''$) etc. The right panel displays the modeled PV cut and spectra from SimLine.}
    \label{fig:SimLine2}
\end{figure*}

We use the 1D radiative transfer code SimLine~\citep{Ossenkopf2001} to model the observed \CII\ emission from a symmetric spherical geometry. SimLine self-consistently solves the excitation problem of any species by taking into account the collisional excitation from the surrounding gas and the radiative interaction throughout the cloud. After solving the excitation problem through an accelerated $\Lambda$-iteration, the line profiles are computed at any desired velocity and spatial resolution. We simulate the properties of the observed \CII{} data at 15$''$ resolution. Collisional excitation in the \hii{} region is assumed to come from electrons in a fully ionized medium. In the outer shells, we assume that collisions are dominated by H$_2$, although a contribution of atomic hydrogen is also possible.

Due to the one-dimensional nature of the model, it can only reproduce angular-invariant properties, ignoring variations in different directions. Therefore, we ignore the SE direction which is heavily affected by the foreground molecular cloud and reproduce the observed radial profiles seen in the NW direction indicated by blue squares in Fig.~\ref{fig:SimLine1}, which shows the general shell-setup of the model. We do not perform an accurate $\chi^2$-fit to the data. This would fail due to the radial asymmetry given by the overabundance of foreground material from the molecular cloud at the peak position relative to the background material and the contribution of the high-velocity wing material that we ignored as discussed in Sect.~\ref{subsec:self}. Instead, we performed a qualitative fit to all significant features by eye, ignoring the wings below $-55\,\kms{}$ and above $-40\,\kms{}$ and the blueshifted material close to the O star position. 

The model consists of a central, fully ionized \HII\ region, followed by a dense layer with high temperature and outer region providing a steep gradient in density, temperature, and expansion velocity ($\propto r^{-\alpha_{s}}$) to the surrounding ambient medium. 

We estimated the approximate extent of the \HII\ region using the 843$\,$MHz emission map at 45$''$ angular resolution (lower left panel in Fig.~\ref{fig:complementary_plots}). First, the \HII\ region must be smaller than the area enclosed by the $8\,\mum$ ring-like feature, represented by the green dashed circle with a radius of 50$''$. Second, the contour at $0.2\,$Jy/beam delineates the level, at which the emission drops to $\sim\,$50\% of its maximum value corresponding to an observed radius of approximately 34$''$. Using this value and a beam of 45$''$, we derive a de-convolved radius of the \HII\ region of $\sim\,$$0.5\,$pc.
We note that this is only an approximation, particularly as we see clear deviations from a circular symmetry. The \HII\ region may be slightly smaller, suggested by the \textit{Spitzer} 8$\,\mum$ image, as a small circular feature is visible in Fig.~\ref{fig:spitzer}, located directly around the exciting O star. Higher angular resolution cm observations are necessary to resolve this uncertainty. 
Beyond the \HII\ region, the model assumes a PDR with a transition from atomic to molecular gas, where the outermost layer provides the transition to the gas from the surrounding ambient molecular cloud.

We setup the shells in the following way (all parameters are summarized in Table~\ref{tab:SimLine}): In the inner \HII{} region (up to $0.5\,$pc), we assume that carbon is singly ionized. 
We cannot exclude that some of the C$^+$ is photo-ionized to C$^{2+}$ caused by the stellar radiation~\citep{Ebagezio2023, Ebagezio2024}, but this effect should be more prominent in more evolved \HII\ regions~\citep{Ebagezio2024}. 
We take a typical temperature of $8000\,$K for the \HII\ region. The density has no impact on the model result as long as it is below a few hundred $\mathrm{cm^{-3}}$. We choose $100\,\mathrm{cm^{-3}}$ as a typical value. 

The following two shells constitute the PDR, in which carbon collides with atomic and molecular hydrogen. For these shells, we adopt an abundance of $\mathrm{X(C^+/H)}=1.2\times10^{-4}$~\citep{SimonDiaz2011}. The first shell (up to $0.95\,$pc) has no velocity, density, or temperature gradient. It expands with the velocity of $2.6\,\kms{}$ as determined in Sect.~\ref{subsec:overview} and has a turbulent velocity dispersion of $v_\mathrm{turb}=2.2\,\kms{}$ to match the \CII\ line width, a temperature of $T=100\,$K reflecting typical PDR conditions, in which the \CII\ line cools efficiently, and a density of $n_{\mathrm{H_2}}=2500\,\mathrm{cm}^{-3}$. 

This matches the density from the \textit{Herschel} column density map at 18$''$ resolution from the HOBYS keyprogram~\citep{Motte2010} presented in~\citet{Liu2017}. The cutout for the bubble is shown in Fig.~\ref{fig:complementary_plots}. Since RCW79 is embedded in the Galactic plane, the column density is overestimated. Following the procedure presented in~\citet{Schneider2015}, that estimates the line-of-sight contamination directly from the \textit{Herschel} maps, we derived a value of $4-6\times 10^{21}\,\mathrm{cm^{-2}}$ for the contaminating column density. 
This is a typical value for massive star-forming regions~\citep{Schneider2022}. 
After subtracting a value of $4\times 10^{21}\,\mathrm{cm^{-2}}$, we calculated densities between 2.3 and $2.9\times10^{3}\,\mathrm{cm^{-3}}$ in an 18$''$ beam for the outermost points of shell 2 and 3, which fits our fit value of $2500\,\mathrm{cm^{-3}}$.

The outer region (up to $1.5\,$pc), representing the transition to the pre-shock material in front of the expanding first shell, is simulated through a power-law decay of all parameters, with a steep exponent of $\alpha_{s}=2$ for density, temperature, and expansion velocity; and a shallower exponent of $\alpha_{s}=1$ for the turbulent velocity, reflecting the indirect driving of turbulence through the expansion. This description should roughly mimic the gas properties and condition in front of a C-shock, where the temperature and density rise steeply.

The result is shown in Fig.~\ref{fig:SimLine2} comparing the PV cut through the model and selected spectra with the correspondingly observed data at the same scale. 
Our SimLine setup with an absorbing layer, including temperature, radial, and turbulent velocity gradients, shows a good agreement to our observed spectra in terms of intensity and line profiles and PV cut. The most evident difference is the reversion of the blue and red peaks for positions $\Delta r = 0$ and 21. 
In an ideal bubble scenario including absorption, a higher redshifted line is expected since the blueshifted gas experiences more absorption along the sightline.
However, the geometry of the C$^+$ bubble is more complex, as seen in Fig.~\ref{fig:SimLine1}. The spherical symmetry of the \HII\ region and the expanding C$^+$ bubble is disturbed by the molecular gas distribution located SE of the O star, which is slightly more blueshifted with respect to the \CII\ emission. 
Therefore, the PDR surfaces of the molecular clumps emit \CII\ efficiently and the increased gas density results in stronger \CII\ emission.

\section{Determination of the total far-infrared flux}
\label{app:SED_fitting}

In order to determine a total FIR flux, we create dust spectral energy distributions (SEDs) using the \textit{Herschel} PACS 70$\,\mum$ and 160$\,\mum$ and SPIRE 250 to 500$\,\mum$ bands. All maps are convolved and re-gridded to the largest common angular resolution of $36''$ with an $8''$ pixel size. Following~\citet{Pabst2022}, 
we fit a gray body given by 
\begin{align}
I_{\lambda} = B(\lambda, T_{d}) \left[1 - \exp{\left(-\tau_{160} \left(\frac{160 \mu m}{\lambda}\right)^\beta\right)}\right]~,
\end{align}
with a fixed emissivity index $\beta=2$ and an optical depth specified at $160\,\mum$. 
A total FIR flux is then determined by integration between 40 to 500$\,\mum$. The choice of fixing $\beta$ biases the SED fit toward shorter wavelengths and warmer dust, but allows comparison with above mentioned studies. We note that the overall effect on the determined total FIR intensity is not strong, and the discussed correlations remain insensitive to this choice. The dominant error in the SED fit is the individual flux uncertainties in the \textit{Herschel} PACS and SPIRE bands, which we assume to be 20\% and 10\%, respectively.

\end{appendix}

%
%


\end{document}